\documentclass[reprint,nofootinbib]{revtex4-1}

\usepackage[bitstream-charter]{mathdesign}
\usepackage{booktabs}
\usepackage[english]{babel}
\usepackage{graphicx,xcolor}
\usepackage{textcomp}
\usepackage{amsmath}
\usepackage{gensymb} 

\renewcommand{\phi}{\ensuremath{\varphi}}

\newcommand{\vidx}[2]{\ensuremath{#1_{\mathrm{#2}}}}
\newcommand{\ang}[1]{#1\textdegree}

\newcommand\blfootnote[1]{%
  \begingroup
  \renewcommand\thefootnote{}\footnote{#1}%
  \addtocounter{footnote}{-1}%
  \endgroup
}
\begin{document} 

    \title[Traceable GISAXS for 25~nm grating pitch measurements]{Traceable GISAXS measurements for pitch determination of a 25 nm self-assembled polymer grating}
    \author{Jan Wernecke}
    \email[contact: ]{jan.wernecke@ptb.de}
    \affiliation{Physikalisch-Technische Bundesanstalt (PTB), Abbestr. 2-12, 10587 Berlin, Germany}
    \author{Michael Krumrey}
    \affiliation{Physikalisch-Technische Bundesanstalt (PTB), Abbestr. 2-12, 10587 Berlin, Germany}
    \author{Armin Hoell}
    \affiliation{Helmholtz-Zentrum Berlin f\"ur Materialien und Energie (HZB), Albert-Einstein-Str. 15, 12489 Berlin, Germany}
    \author{R. Joseph Kline}
    \author{Hung-kung Liu}
    \author{Wen-li Wu}
    \affiliation{National Institute of Standards and Technology (NIST), Gaithersburg, Maryland 20899, USA}

    \begin{abstract}
        The feature sizes of only a few nanometers in modern nanotechnology and next-generation microelectronics continually increase the demand for suitable nanometrology tools.
        Grazing incidence small-angle X-ray scattering (GISAXS) is a versatile technique to measure lateral and vertical sizes in the nm-range, but the traceability of the obtained parameters, which is a prerequisite for any metrological measurement, has not been demonstrated so far.
        In this work, the first traceable GISAXS measurements, demonstrated with a self-assembled block copolymer grating structure with a nominal pitch of 25~nm, are reported.
        The different uncertainty contributions to the obtained pitch value of 24.83(9)~nm are discussed individually.
        The main uncertainty contribution results from the sample-detector distance and the pixel size measurement, whereas the intrinsic asymmetry of the scattering features is of minor relevance for the investigated grating structure.
        The uncertainty analysis provides a basis for the evaluation of the uncertainty of GISAXS data in a more general context, for example in numerical data modeling.
    \end{abstract}

    \maketitle

\section{Introduction}
Modern nanotechnology \cite{nanotech} offers a wide range of prospective applications, for example in materials science, electronics, communications, or drug delivery.
One of the strong driving forces is the microelectronics industry, together with the compliance with ``Moore's law'' of doubling the number of transistors on a chip every 18 months. 
New materials and ever decreasing structure sizes down to the atomic scale are needed for future devices because traditional silicon MOSFET technology is reaching its limits \cite{vogel_tech_2007,muller_sound_2005}.
Next-generation photolithography tools in the extreme UV-wavelength regime (EUV lithography) \cite{wagner2010euv} and directed self-assembly of block copolymer (BCP) thin films \cite{leibler1980bcptheory,darling2007directing,albert2010self,liu2013chemical,chang_design_2014,bates2014} are promising techniques for producing a wide variety of structures with great accuracy and dimensions down to several nanometers \cite{hamley2003nanofab,saavedra_hybrid_2010}.
Not only microelectronics, but also other emerging fields like organic photovoltaics \cite{brabec2014organic,schaffer_direct_2013}, nanotemplating \cite{pechkova2005,nie_patterning_2008}, or surface functionalization \cite{rosler2012advanced,husemann2000} make increasing use of structuring in the nanometer range by thin film deposition, processing, and self-assembly techniques.

What all these different applications have in common is the need for suitable metrology tools to measure surface and subsurface structural parameters with sufficient accuracy, which is the field of dimensional nanometrology \cite{leach2011european,co-nanomet-foresight}.
Although there is a large variety of different techniques available, ranging from direct methods like critical-dimension electron microscopy (CD-SEM) \cite{bunday2003cd,villarrubia2005scanning} and atomic force microscopy (AFM) \cite{misumi2003uncertainty,yacoot2011recent} to indirect methods like X-ray small-angle scattering (SAXS) and grazing incidence SAXS (GISAXS) \cite{krumrey2011synchrotron,wernecke2012,hofmann2009}, critical dimension SAXS (CD-SAXS) \cite{jones2003,hu2004,Sunday2014}, or extreme UV (EUV) scatterometry \cite{gross2009profile,perlich2004characterization}, all of them have very specific advantages and drawbacks.
Moreover, only a few of them are traceable, that is, related to the International System of Units (SI system) by an unbroken chain of comparisons with known uncertainty, which is ultimately required in order to associate uncertainty values with any measured quantity.

A versatile technique that provides access to lateral and vertical structure dimensions in the nanometer range in a fast, non-destructive, non-contact, in-situ capable way is GISAXS \cite{levine1989,renaud2009}.
It is a now widely used synchrotron X-ray technique, especially for structured polymer films \cite{pmb2003,lee2005,stamm2008,wang2011}, that offers access also to buried structures and depth-resolved measurements \cite{okuda2011near,hoydalsvik2010} in thin films and layer systems.
Moreover, the tunable photon energy of a synchrotron beamline provides access to GISAXS measurements of element-specific spatial distributions by anomalous scattering and to contrast variation techniques.
In terms of nanometrology, GISAXS measurements on gratings towards a traceable grating pitch determination have been reported \cite{wernecke2012}.

In this work, we report the first traceable GISAXS measurements, demonstrated on a self-assembled BCP thin film grating with a nominal pitch of 25~nm.
The sample system has been chosen because it offers GISAXS data analysis of the structure factor alone, thus primarily yielding the experimental uncertainty contribution.
Nanostructured BCP thin films are not just academic model systems, but are highly relevant and in use in new technology fields such as, for example, block copolymer lithography \cite{bates2014,son2013sub} and in organic photovoltaics \cite{brabec2014organic}.

\section{Theory}
\label{sec:theory}

\begin{figure}
    \includegraphics{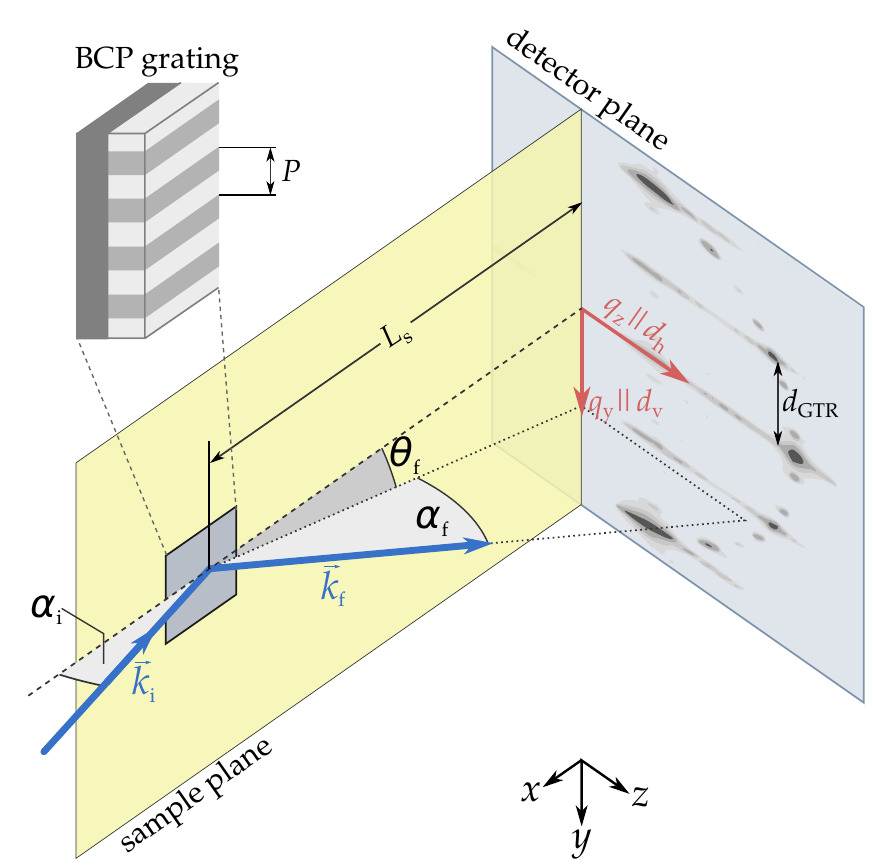}
    \caption{GISAXS scattering geometry.
        The coordinates $(x,y,z)$ denote the laboratory coordinate system, $(\vidx{q}{x},\vidx{q}{y},\vidx{q}{z})$ the reciprocal space coordinates, and $(\vidx{d}{h},\vidx{d}{v})$ the horizontal and vertical detector coordinates.
        The enlargement shows an illustration of the self-assembled block copolymer grating with a nominal pitch $P$ of 25~nm.
    }
    \label{fig:gisaxs-geo}
\end{figure}

GISAXS is a technique that probes the reciprocal or momentum transfer space.
The momentum transfer is defined as the difference between incident X-ray beam $\vidx{\vec{k}}{i}$ and elastically scattered beam $\vidx{\vec{k}}{f}$ by
$\vec{q} = \vidx{\vec{k}}{i} - \vidx{\vec{k}}{f} = \left(\vidx{q}{x},\vidx{q}{y},\vidx{q}{z}\right)^T$.
The components of $\vec{q}$ are related to the angles of incidence and scattering (Figure~\ref{fig:gisaxs-geo}) according to
\begin{equation}
    \begin{split}
        \vidx{q}{x} &= k \, \left(\cos\vidx{\theta}{f} \cos\vidx{\alpha}{f} - \cos\vidx{\alpha}{i} \right) \ , \\
        \vidx{q}{y} &= k \, \left(\sin\vidx{\theta}{f} \cos\vidx{\alpha}{f} \right) \ , \\
        \vidx{q}{z} &= k \, \left(\sin\vidx{\alpha}{i} + \sin\vidx{\alpha}{f} \right) \ ,
    \end{split}
    \label{eq:q-gisaxs}
\end{equation}
with the wave vector $k = 2\pi / \lambda$. 
In GISAXS geometry, the grazing incidence angle \vidx{\alpha}{i} is kept at a fixed value around or below \ang{1} and the 2-dimensional X-ray detector is placed several meters away from the sample.
Thus, the detector surface is almost fully congruent with the \vidx{q}{y}-\vidx{q}{z} plane of scattering \cite{renaud2009}.
X-rays are scattered due to changes of the complex refractive index $n = (1-\delta) + i\beta$ in the penetrated volume.
A critical angle $\vidx{\alpha}{c} = \sqrt{2\delta}$ (valid for negligible absorption, which means $\delta \gg \beta$) exists below which total reflection occurs at the interface to a denser material.
A significant amount of multiple scattering-reflection events contributes to the scattering pattern especially in the case of $\vidx{\alpha}{i} \approxeq \vidx{\alpha}{c}$.
In general, such multiple scattering effects are accounted for by using semi-kinematic theories like the Distorted Wave Born Approximation (DWBA) \cite{sinha1988,holy1994,salditt1995}, dynamic multilayer theory \cite{wu1993,wu1994}, or rigid vector theory.
Due to the phase problem, that is the loss of phase information in the recorded scattering image, this involves the selection of appropriate models for the form factor, the structure factor, and the distribution function to simulate the experimental GISAXS data \cite{isgisaxs,fitgisaxs}.

The particular case of GISAXS on surface gratings has been extensively investigated \cite{jergel1999,mikulik2001,yan2006,yan2007,hofmann2009} and shall be only briefly reviewed.
In the perfectly parallel alignment of the projected incident beam and grating lines, the GISAXS pattern consists of sharp spots aligned on a semicircle, equidistantly spaced along \vidx{q}{y}.
The pattern can be understood within the framework of reciprocal space construction \cite{mikulik2001}.
The semicircular shape is a consequence of the condition of elastic scattering, $|\vidx{\vec{k}}{i}| = |\vidx{\vec{k}}{f}|$ with a radius of $2\pi / \lambda$ (Ewald sphere).
The reciprocal space representation of a (perfect) line grating with a pitch (period length) $P$ consists of rods that are extended in the \vidx{q}{z} direction with a periodicity of $\Delta\vidx{q}{y} = 2\pi / P$ in the direction of \vidx{q}{y}, the so called grating truncation rods (GTR).
The scattering pattern on the detector is the intersection of the Ewald sphere with the reciprocal space representation of the grating, hence the appearance of maxima equidistantly aligned on a semicircle.
It is known that the GISAXS pattern is very sensitive to smallest azimuthal rotations $\phi$ away from the parallel orientation.
Deviations of \ang{0.002} can already be observed by a visual distortion of the scattering image.
Due to geometrical reasons \cite{yan2006}, the intersection of Ewald sphere and GTRs bends up in one direction and the GTR scattering spots move along $\pm\vidx{q}{y}$.

The parameter that can be extracted directly from such data without any semi-kinematic or dynamic modeling is the grating pitch $P$, which is termed as ``direct analysis'' throughout this article.
This is essential for the evaluation of the experimental uncertainties of parameters determined with GISAXS as it avoids the complex analysis by numeric simulation, which would induce unknown additional uncertainties by model assumptions.
Thus, for small angles, the grating pitch $P$ can be determined in this simple picture by the grating equation, which is equivalent to the structure factor of a grating in reciprocal space \cite{koch_handbook_1983,yan2007}.
\begin{equation}
    P = \frac{m \lambda}{2 \cos\phi \sin(\vidx{\theta}{f})_m}.
    \label{eq:bragg}
\end{equation}
By replacing the photon wavelength $\lambda$ and the azimuthal scattering angle $(\vidx{\theta}{f})_m$ of the $m$-th order GTR by the actual experimental input parameters photon energy $\vidx{E}{ph}=\tfrac{h\,c}{e\, \lambda}$ ($h$~--~Planck constant; $c$~--~speed of light; $e$~--~elementary charge), sample-detector distance \vidx{L}{s}, detector pixel size \vidx{L}{px}, and GTR distance \vidx{d}{GTR} in detector pixel units, we obtain the equation
\begin{equation}
    P = \frac{m\, h\, c}{e\, \vidx{E}{ph}} \ \frac{1}{2\cos\phi \ \frac{\vidx{d}{GTR}\, \vidx{L}{px}}{\vidx{L}{s}}} .
    \label{eq:bragg-exp}
\end{equation}

Before the analysis is carried out, the applied assumptions of the approach need to be discussed.
Following the Born approximation and the convolution theorem, the scattered intensity can be written as a product of the form factor, which describes the average object shape, and the structure factor, which describes the arrangement of objects \cite{naudon_grazing-incidence_2000}.
The structure factor of a grating with a pitch $P$ gives rise to the discussed GTRs, separated by $\delta\vidx{q}{y} = 2\pi/P$.
The form factor requires a suitable model for the scattering objects, for example spheres, cylinders or more complex shapes.
This is problematic for traceability, because it is impossible to evaluate the `correctness' of the model selection itself by an uncertainty analysis within the model.
The intensity of the GTR peaks depends on the form factor as well as on the structure factor \cite{panduro_using_2014}, but their positions are solely governed by the structure factor, hence, independent from the actual cross-sectional line shape.
Thus, by restricting the analysis to the peak positions to determine the grating pitch $P$, additional non-traceable assumptions about the form factor of the grating lines are avoided.
The underlying assumption of this approach is the translational symmetry of the grating within the illuminated area.

\section{Experimental Setup}
    \blfootnote{Certain commercial equipment, instruments, or materials are identified in this report in order to specify the experimental procedure adequately.
        Such identification is not intended to imply recommendation or endorsement by the National Institute of Standards and Technology, nor is it intended to imply that the materials or equipment identified are necessarily the best available for the purpose.}

    \subsection{Sample preparation}
    Directed self assembly of a lamellar phase of polystyrene-poly(methyl methacrylate) block copolymer (\mbox{PS-\textit{b}-PMMA}) on a 25~mm $\times$ 25~mm silicon wafer was carried out via a frequency quadrupling process by the following procedure \cite{cheng2010fabri}:
    A line grating template with a nominal pitch of 100~nm was prepared with 193~nm water immersion lithography, followed by the deposition of a neutralization layer and a lift-off process.
    The PMMA blocks were etched off and the resultant PS line gratings possess a nominal  pitch of 25~nm (sample courtesy of J. Y. Cheng, IBM Almaden Research Center, CA). 
    The silicon substrate was fully coated with the polymer film.
    Details of the sample preparation can be found in the literature \cite{cheng2010fabri}.

    \subsection{Instrumentation and Traceability}
    \label{sub:instrum}
    All GISAXS measurements were performed at the four-crystal monochromator (FCM) beamline \cite{krumrey2001} of the PTB at the synchrotron radiation facility BESSY~II of the Helmholtz-Zentrum Berlin (HZB) \cite{ptb-quarterc}.
    The beamline covers a photon energy range of 1.75~keV to 10~keV with a beam size of $0.3\ \mathrm{mm} \times 0.3\ \mathrm{mm}$.
    Traceability of the energy scale has been established by relating the photon energy \vidx{E}{ph} to the lattice constant of silicon via back-reflection from a silicon single crystal.
    For photon energies above 2.1~keV, four Si(111) crystals are used in the monochromator.
    The resolving power is above $10^4$, which yields an uncertainty of $u(\vidx{E}{ph}=10\,000\ \mathrm{eV}) = 1\ \mathrm{eV}$ for the performed measurements \cite{krumrey2001}.
    A sample chamber equipped with six axes for sample movement is attached to the FCM beamline \cite{fuchs-newref}.
    For SAXS and GISAXS measurements, the SAXS instrument of the HZB \cite{krumrey2011synchrotron,hoell-saxs-patent} is installed behind the sample chamber, which provides the positioning of the 2D detector.
    The CCD-based detector (MarCCD, sensitive area diameter of $165\ \mathrm{mm}$) is installed on a movable sledge and connected to an edge-welded bellow to allow for any sample-to-detector distance between $2.3\ \mathrm{m}$ and about $4.5\ \mathrm{m}$, and a vertical tilt angle up to 3\textdegree\ without breaking the vacuum.
    The vertical movement is realized by two translation axes.
    Both vertical axes as well as the horizontal distance variation axis are equipped with optical encoders (calibrated Heidenhain LC~182, ST~3008, and MT25B) which measure the displacement with an accuracy of 0.001~mm (ST~3008 and MT25B) and 0.005~mm (LC~182), respectively.
    These encoders establish the traceability of the detector displacement along these axes.
    The rear end (detector side) of the bellow holds a movable beamstop to block the intense transmitted or specularly reflected fraction of the beam.
    Figure~\ref{fig:gisaxs-geo} shows the orientation and notation of the various coordinate systems (laboratory, detector, reciprocal space).
    Note that the sample was mounted in an upright position to determine the GTR distance and the pixel size along the traceable vertical direction.
    Consequently, the \vidx{q}{z} axis is parallel to the horizontal detector coordinate \vidx{d}{h} to comply with the common orientation convention of the \vidx{q}{x,y,z} coordinates.

    Traceability of the pitch determination with GISAXS is established by tracing all input quantities of equation (\ref{eq:bragg-exp}).
    Hence, the uncertainties of sample-detector distance, azimuthal angular misalignment from parallel orientation, pixel size, and GTR positions are discussed and evaluated one by one in the following section.
    Then, the grating pitch is determined in a traceable way with these input parameters and corresponding uncertainty contributions according to the \emph{Guide to the Expression of Uncertainty in Measurement}\footnote{available at \texttt{http://www.bipm.org/en/publications/guides/gum.html}.} (GUM) \cite{gum}.

\section{Results and Discussion}

    \subsection{Sample-detector distance}
    \label{sub:sample-det-dist}

    \begin{figure}
        \includegraphics{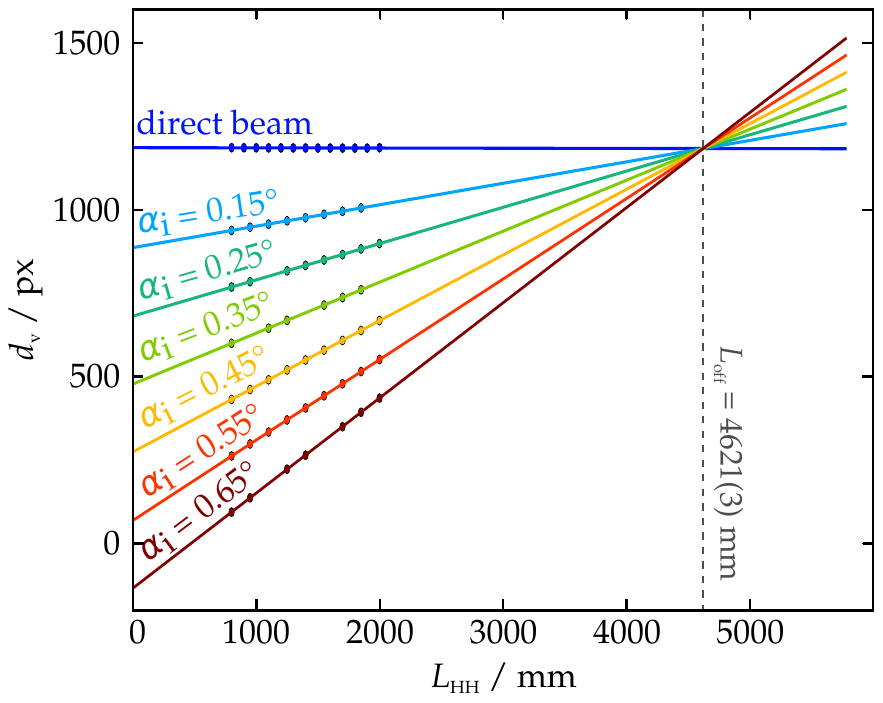}
        \caption{The vertical reflection spot positions at different relative detector positions and different incidence angles \vidx{\alpha}{i} were simultaneously fitted by linear functions with a common intersection point at $\vidx{L}{HH} = \vidx{L}{off}$.}
        \label{fig:ls}
    \end{figure}

    The HZB SAXS instrument is equipped with a Heidenhain encoder to measure the relative position of the detector along the $x$-axis, \vidx{L}{HH}, with \textmu m precision, but the entire setup can be moved with respect to the beamline.
    Thus, it is necessary to determine the offset distance \vidx{L}{off} between the sample and the position of the instrument to obtain the sample-detector distance \vidx{L}{s} by $\vidx{L}{s} = \vidx{L}{off} - \vidx{L}{HH}$ (\vidx{L}{HH} value decreases with increasing sample-detector distance).
    This is achieved by triangulation with the incoming X-ray beam.
    The incident beam is specularly reflected from the sample surface and impinges on the detector at a designated position $(\vidx{d}{h,spec},\vidx{d}{v,spec})$.
    The vertical spot position \vidx{d}{v,spec} is determined for six different incidence angles $\vidx{\alpha}{i}$ ranging from 0.15\textdegree\ to 0.65\textdegree\ and at 13 relative detector positions to cover a range of \vidx{L}{HH} of 1.6~m (Figure~\ref{fig:ls}).
    Then, the data are grouped by incidence angle and simultaneously fitted with linear functions that have individual parameters for the slope and a common intersection point $(\vidx{L}{off},\vidx{d}{v,isp})$.
    The intersection point position \vidx{L}{off} indicates the offset distance under the assumption that the position of reflection on the sample surface is independent of \vidx{\alpha}{i} and \vidx{L}{HH}, which was verified by visual observation.
    The resulting value \vidx{L}{off} and its standard deviation of $(4621 \pm 3)$~mm and the relative detector position during the GISAXS measurement of $\vidx{L}{HH} = (1797.704 \pm 0.003)~\mathrm{mm}$ yield a sample-detector distance \vidx{L}{s} of
    \begin{equation}
        \vidx{L}{s} = \left( 2823 \pm 3 \right)\ \mathrm{mm} .
        \label{eq:Ls}
    \end{equation}

    It has to be noted that the geometric footprint of the beam along the incidence direction of $\tfrac{V}{\tan\vidx{\alpha}{i}}$ with a vertical beam size of $V = 0.3$~mm is larger than the sample length of 25~mm.
    Hence, \vidx{L}{s} is more precisely the distance between the sample rotation axis \vidx{\alpha}{i} and the detector surface.
    Throughout this manuscript, ``sample-detector distance'' is still used as it is the term that is the most familiar to most X-ray scattering users, but it should be understood according to the above definition.

    Care has to be taken that the sample height has to be properly aligned to the half beam position at all times as small deviations can already change the sample-detector distance by several millimeters.
    During the measurements, this can be checked with a photodiode at the direct beam position behind the sample, which should show half the signal of the full beam diode current at an incidence angle of \ang{0}.
    Additionally, the position of the specularly reflected beam on the detector should be monitored during the measurements.

    \subsection{Misalignment from parallel orientation}
    \label{sub:misalignment}

    \begin{figure}
        \includegraphics{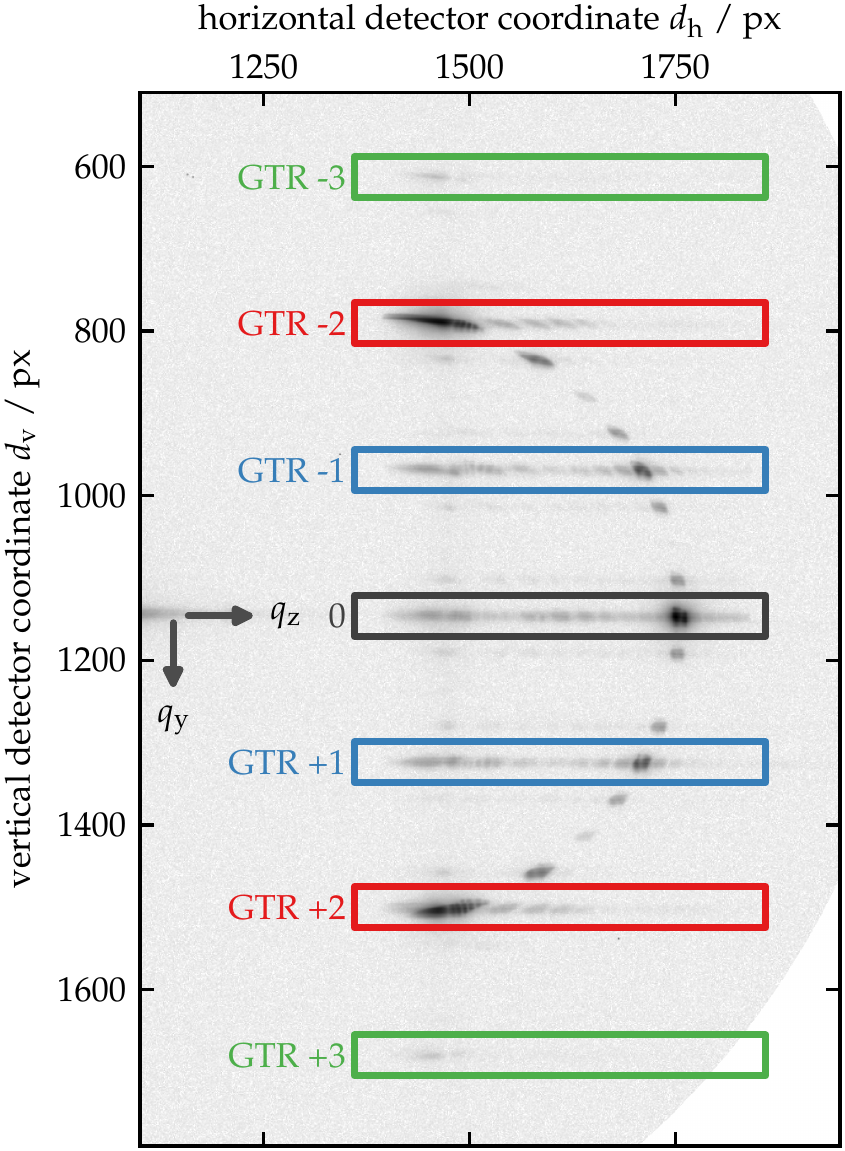}
        \caption{GISAXS pattern in parallel alignment of lines and incident beam.
        The boxes are so defined as to enclose the grating truncation rods (GTR) (-3,\ldots,+3) and the specular axis (labeled as '0') without truncating the rod or including additional scattering features.
        Within each box, the center-of-mass position of the GTR is determined.}
        \label{fig:gisaxs-ov}
    \end{figure}

    \begin{figure}
        \includegraphics{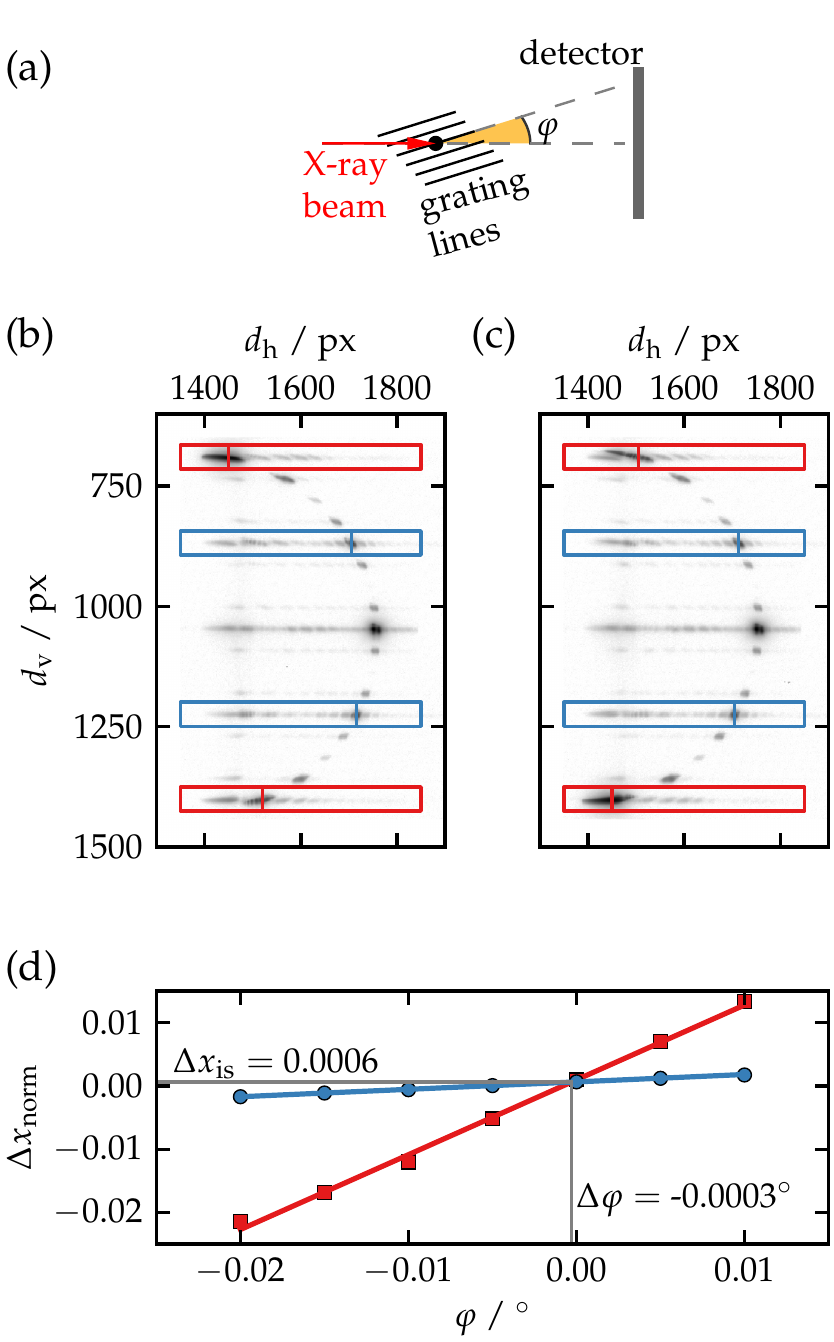}
        \caption{Evaluating the misalignment from parallel orientation of incident beam and grating lines:
            (a) Definition of the azimuthal tilt angle $\phi$.
            (b,c) Smallest and largest tilt angle \ang{-0.020} and \ang{0.010} (step size \ang{0.005}), respectively.
            The boxes of GTRs $\pm 2$ (red) and GTRs $\pm 1$ (blue), as defined in Fig.~\ref{fig:gisaxs-ov}, indicate the regions used to determine the center-of-mass positions of the GTRs (vertical line inside each box).
            (d) Plots of horizontal top/bottom spot pair distances divided by their sum, $\Delta\vidx{x}{norm} = \frac{\vidx{d}{h}\text{(top)} - \vidx{d}{h}\text{(bottom)}}{\vidx{d}{h}\text{(top)} + \vidx{d}{h}\text{(bottom)}}$, of GTRs $\pm 2$ (red) and GTRs $\pm 1$ (blue) as a function of $\phi$.
            The intersection $\Delta\phi$ represents the misalignment from parallel orientation of beam and lines.}
        \label{fig:misalign}
    \end{figure}

    A typical GISAXS pattern of parallel orientation of incident beam and grating lines is displayed in Figure~\ref{fig:gisaxs-ov}.
    In GISAXS geometry, the symmetry of the pattern is very sensitive to smallest deviations from parallel orientation \cite{mikulik2001,yan2007,wernecke2012}.
    Since a misalignment $\phi \ne 0\degree$ (Figure~\ref{fig:misalign}a) would directly result in a distortion of the pitch by a factor of $\cos(\phi)$, it needs to be quantified for a traceable pitch determination.
    A series of 7 GISAXS images at a fixed photon energy of 10\,000~eV and a fixed incidence angle of $\vidx{\alpha}{i} = 0.57\degree$ has
    been recorded for different azimuthal rotation angles $\phi = (-0.020\degree \ldots 0.010\degree)$ around the most parallel position at $\phi = 0\degree$ (Figure~\ref{fig:misalign}a-c).

    The misalignment is evaluated by the following objective procedure:
    Boxes were defined to surround each GTR as well as the specular axis on one of the GISAXS images (Figure~\ref{fig:gisaxs-ov}).
    The placement and dimensioning of the boxes follow two rules: (i) The rod and diffraction spot must not be truncated and (ii) additional scattering features must not be included in any image of the rotation series.
    After this initial placement, the boxes remain fixed in size and position for every rotation angle.
    Within each box, a filter is applied to cut off the lowest $10\ \%$ of counts inside the box to remove the detector background, cosmic radiation, and other weak irrelevant scattering features.
    It was checked beforehand to ensure that the effect of the filter is negligible for the determined GTR position (variation within 0.5~pixel for a filter threshold between $5\ \%$ and $95\ \%$).
    Then, the center-of-mass (COM) position $(\vidx{d}{h,COM}, \vidx{d}{v,COM})$ of the GTR is determined for each evaluated box to find the diffraction peak position.

    The misalignment was analyzed by the diffraction peak positions \vidx{d}{h,COM} of the GTR pairs $\pm 2$ (red boxes in Figs.~\ref{fig:gisaxs-ov} and \ref{fig:misalign}b,c) and $\pm 1$ (blue boxes) determined throughout the rotation series.
    Parallel alignment of grating lines and incident beam is reached if \vidx{d}{h,COM} of both spots of a GTR pair ($\pm 2$ and $\pm 1$, respectively) is equal, which means $\vidx{d}{h,COM\ top} - \vidx{d}{h,COM\ bottom} = 0$.
    In order to compare both GTR pairs, the difference is normalized by the sum of both positions, $\Delta \vidx{x}{norm} = \frac{\vidx{d}{h,COM\ top} - \vidx{d}{h,COM\ bottom}}{\vidx{d}{h,COM\ top} + \vidx{d}{h,COM\ bottom}}$.
    The intersection of \vidx{\Delta x}{norm} as a function of $\phi$ of GTR pair $\pm 2$ and $\pm 1$ yields the misalignment $\Delta\phi$ as well as the deviation from perpendicular orientation of the sample surface and the detector \vidx{\Delta x}{is} (Figure~\ref{fig:misalign}d).
    The obtained values of $\Delta\phi = -0.0003\degree$ and $\vidx{\Delta x}{is} = 0.0006$ illustrate that the detector is well aligned in terms of incident beam and grating lines as well as \vidx{q}{z} being parallel to \vidx{d}{h} at $\phi = 0\degree$.
    Since $\cos\left(-0.0003\degree\right)$ deviates by less than $10^{-10}$ from unity, the term can be completely neglected for the calculation of the pitch $P$ and its uncertainty.

    \subsection{Pixel size}
    \label{sub:px-size}

    \begin{table}
        \caption{Vertical detector displacement positions, nominal values and values measured with Heidenhain encoders attached to both lifting axes.}
        \label{tab:displ}
        \begin{tabular}{llllll}
            \toprule
            nominal value / mm:  & 0     &  2     & 4     &  6     & 8     \\
            measured value / mm: & 0.000 &  2.002 & 4.009 &  6.010 & 8.007 \\
            \bottomrule
        \end{tabular}
    \end{table}

    For pixel size determination, the relation between the absolute length scale measurement and the corresponding number of pixels on the detector needs to be established \cite{invac-pilatus}.
    For that purpose, the detector was shifted vertically in steps of 2~mm up to a total displacement of 8~mm (nominal values).
    GISAXS images (similar to Figure~\ref{fig:gisaxs-ov}) at $\vidx{E}{ph} = 10\,000$~eV and $\vidx{\alpha}{i} = 0.57\degree$ were recorded at each position and the real displacement was measured with the Heidenhain encoders attached to both lifting axes (Table~\ref{tab:displ}). 
    The beam footprint at this incidence angle is 30~mm, which is longer than the sample length.
    The uniformity of the GISAXS patterns and the GTR positions within 1~pixel has been verified by comparing images recorded at different sample stage positions along both directions parallel to the sample surface.
    Moreover, a change of horizontal beam width from 0.3~mm to 1.0~mm did not cause any detectable changes of the GTR positions on the detector.
    The measured displacement values are the mean of both encoder readings, which had a relative standard deviation of $\le 0.2\ \%$ in all measurements.

    \begin{figure}
        \includegraphics{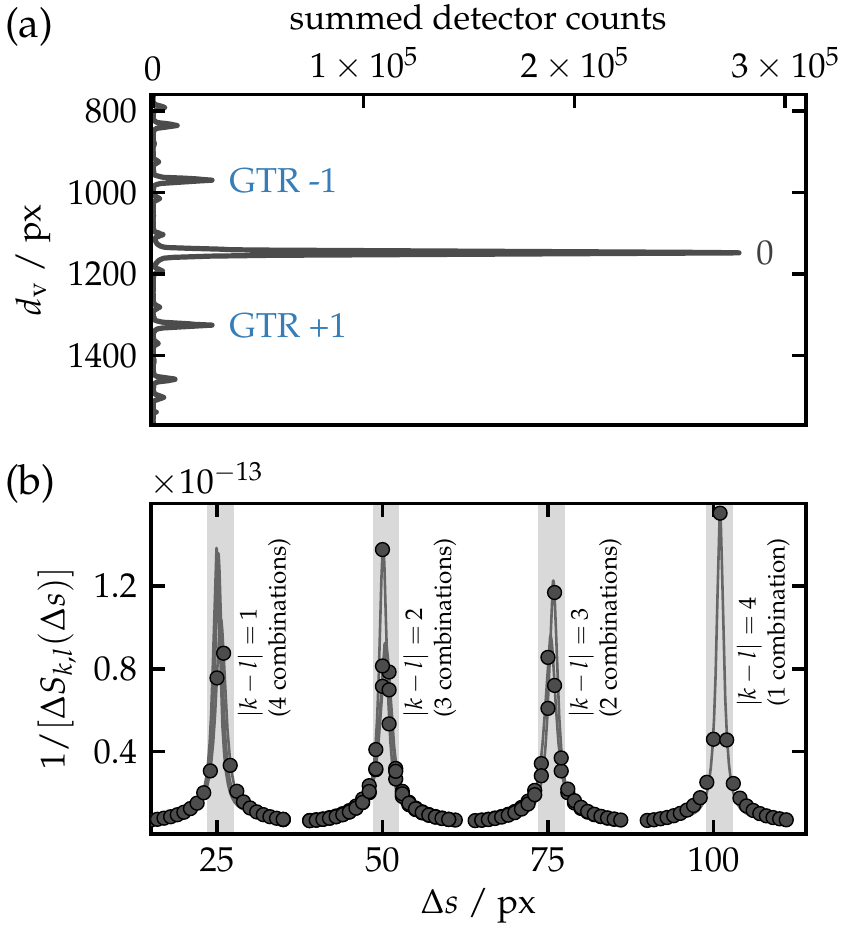}
        \caption{Pixel size determination:
                (a) Exemplary vertical profile $S_k(\vidx{d}{v})$ of the vertical displacement series.
                Each point of the profile represents the sum of counts within the \vidx{d}{h} range $[1530,1820]$ (see Fig.~\ref{fig:gisaxs-ov} for orientation).
                (b) Reciprocal of the difference profile $\left[\Delta S_{k,l}(\Delta s)\right]$ between the profile $S_k(\vidx{d}{v})$ and the shifted profile $S_l(\vidx{d}{v}+\Delta s)$ for all combinations of vertical positions $k$ and $l$ (symbols).
                The reciprocal values of the $\Delta S_{k,l}$ profiles are used in order to fit Lorentzian functions (lines) to the peaks for sub-pixel resolution of the offset peak positions.
                }
        \label{fig:px-profile}
    \end{figure}

    From each image, the relevant section containing the scattering pattern is extracted (\vidx{d}{h} range is $[1530,1820]$; full \vidx{d}{v} range).
    The row-wise sum (i.e., along \vidx{d}{h}) of every subimage is calculated, which results in five profiles $S_{k}(\vidx{d}{h})$, one for each vertical detector position $k=(1,\ldots,5)$ (Fig.~\ref{fig:px-profile}a).
    For each pair $\left[S_k(\vidx{d}{v}),S_l(\vidx{d}{v})\right]$ (i.e., 10 combinations), one of the two profiles is shifted with respect to the other along the
    \vidx{d}{v} axis by $\Delta s$; the absolute difference between both profiles is calculated and summed over \vidx{d}{v}, i.e. 
    $\Delta S_{k,l}(\Delta s) = \sum_{\vidx{d}{v}} \left| S_k(\vidx{d}{v} + \Delta s) - S_l(\vidx{d}{v}) \right|$.
    For a more convenient analysis, the reciprocal of $\Delta S_{k,l}(\Delta s)$ was calculated so that it yields maxima \vidx{\Delta s}{min,k,l} at the offsets between the two vertical positions $k$ and $l$ (Fig.~\ref{fig:px-profile}b).
    The peak positions can be determined with sub-pixel accuracy by fitting Lorentzian functions, $f(x) = \frac{A_0 / \pi}{1 + \left( \frac{x - x_0}{{\sigma}} \right)^2}  + A_1$.
    The fitted position of a peak $x_0$ represents the position of the least squares minimum; the amplitude $A_0$ is used as a weighting factor.
    The offset positions \vidx{\Delta s}{min,k,l} of each pair are associated with the absolute length measurement of the corresponding vertical displacement \vidx{v}{k,l}.
    The data points are fitted by a linear function; each point is weighed by the peak amplitude $A_0$.
    The slope of the fit function yields the pixel size \vidx{L}{px}; the square root of the fit variance $\sigma^2$ defines the uncertainty,
    \begin{equation}
        \vidx{L}{px} = \left( 79.2 \pm 0.2 \right)\ \text{\textmu m} .
        \label{eq:lpx}
    \end{equation}
    There is still potential for a further reduction of the uncertainty by a larger dataset with a wider vertical displacement range, however, this was not possible at the time of the measurements due to technical constraints.

    \subsection{GTR asymmetry}
    \label{sub:gtr-assym}

    \begin{figure}
        \includegraphics{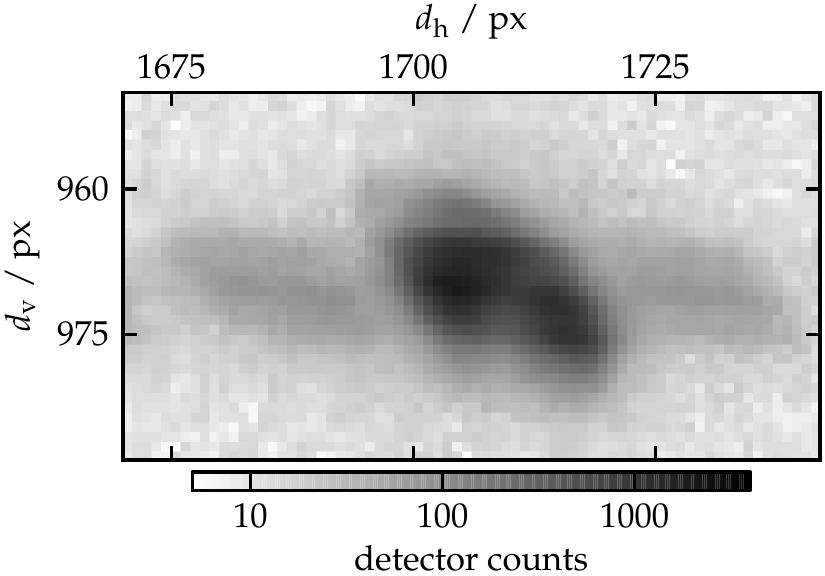}
        \caption{Asymmetric shape of a GISAXS GTR maximum.}
        \label{fig:asym-gisaxs}
    \end{figure}

    Before the positions of the GTRs are determined, a detail needs to be discussed.
    A close examination of the diffraction lines, especially the ones further away from the central line, reveals their lack of symmetry (Figure~\ref{fig:asym-gisaxs}).
    The intensity distribution and symmetry of the intersection of GTRs with the Ewald sphere in grazing incidence geometry is also governed by the form factor of the cross-sectional line profile, see the discussion at the end of section~\ref{sec:theory}.
    Only the structure factor is evaluated by determining the GTR positions in order to maintain traceability, hence, the asymmetry is accounted for, but not quantitatively evaluated by a form factor model.
    Fitting of the form factor and establishing traceability of the shape is a separate research endeavour which is beyond the scope of this work.
    The same applies for the additional scattering features that can be seen in Figures~\ref{fig:gisaxs-ov} and~\ref{fig:misalign}b.
    These are caused by superstructures within the polymer film and can be disregarded for the pitch determination.
    It should be also noted that the observed asymmetry of the diffraction intensity is not caused by an asymmetric incident beam; the incident beam shape was found to be symmetric in a short exposure time image of the attenuated direct beam.

    The asymmetry of the GTRs results in a slight variation of the distance of a GTR from the specular axis, depending on the GTR order.
    This can be assessed by calculating the pitches $P_i$ from the individual GTR distances by
    \begin{equation}
        \begin{split}
            P_i &= \frac{h c}{e \vidx{E}{ph}} \left[2 \cos\Delta\phi \ \left(\frac{\vidx{L}{px}\ \vidx{d}{GTR\,\mathit{i}}}{\vidx{L}{s}\ \left|i\right|}\right) \right]^{-1},\\
            & \mathrm{with}\quad i \in \{-3,-2,-1,+1,+2,+3\}, \\
            & \mathrm{and}\quad P = \frac{1}{N_i}\, \sum_{i} P_i,
        \end{split}
        \label{eq:P2}
    \end{equation}
    instead of averaging the center-of-mass positions of the GTRs and using equation (\ref{eq:bragg-exp}).
    A possibility to quantitatively evaluate the spot asymmetry is to record scattering images at different azimuthal angles $\phi$ around the parallel alignment in small steps (for example, \ang{0.002}).
    In this way, the GTR intersects the Ewald sphere at various distances from the GTR center (most pronounced asymmetry close to the sample horizon).
    Consequently, the peak intensity of the spot on the detector varies as a function of $\phi$.
    The position of maximum intensity yields the center position of the GTR.
    While this highlights the importance and the arising complexity of defining the `right' center position of the diffraction maxima, it should be noted that the asymmetry corrections are usually minor.
    In the present case, as well as in most other cases of gratings with well-defined line structures, the asymmetry correction was insignificant, that is, well within the pitch uncertainty.
    We calculated $P$ with both equations and found a relative deviation of $2\times 10^{-5}$, which is two orders of magnitude below the relative uncertainty of the pitch uncertainty as shown in Table~\ref{tab:uncertainty}.

    \subsection{GTR distances}
    \label{sub:gtr-pos}
    The goal is to calculate the pitch $P$ from the distances of the GTRs from the corresponding specular axis (0).
    The distance is determined by approximating the GTR and the specular axis, each with a linear function and calculation of the average distance of these two lines.
    We evaluated the GTR distances in each of the five images of the vertical displacement series (the same series that was used for the pixel size determination, section~\ref{sub:px-size}), the index $k$ denotes the GISAXS image, $k = (1,\ldots,5)$.
    As before, the index $i$ denotes the GTR order, $i \in \{-3,-2,-1,+1,+2,+3\}$.

    Boxes around the GTRs and the specular axis were defined on each image as illustrated by Figure~\ref{fig:gisaxs-ov} and in the same way as described in section~\ref{sub:misalignment}.
    As before, a $10\ \%$-threshold filter was applied to remove irrelevant weak scattering features and background counts.
    Within each box, a sub-box with full height and a width of 5~px was moved horizontally along \vidx{d}{h} of the GTR box in steps of 5~px.
    At every position $j$ of the sub-box, the vertical center-of-mass position inside the sub-box $\vidx{d}{v,COM}(j)$ is determined.
    In this way, a trace of the GTR (or specular axis) is created that follows the highest intensity along the rod.
    The trace is then approximated by a linear function $\vidx{d}{v}(j) = m\, j\ +\ n$ (with $j = (1,2,\ldots,N_j)$).
    The distance $(\vidx{d}{GTR\ \mathit{i}})_k$ of the $i$-th GTR from the specular axis (\vidx{m}{spec}, \vidx{n}{spec}) of GISAXS image $k$ is determined by the arithmetic mean
    \begin{equation}
        (\vidx{d}{GTR\ \mathit{i}})_k = \frac{1}{N_j} \left(\sum_{j=1}^{N_j} (m_i - \vidx{m}{spec})\, j + (n_i - \vidx{n}{spec})\right) \quad ,
        \label{eq:dgtr-per-img}
    \end{equation}
    the corresponding square root of the variance $(\sigma^2_i)_k$ is taken as the uncertainty of $(\vidx{d}{GTR\ \mathit{i}})_k$.
    The resulting five values $(\vidx{d}{GTR\ \mathit{i}})_k$ and $(\sigma^2_i)_k$ for each GTR $i$ are then used to calculate the weighted arithmetic mean \vidx{d}{GTR\ \mathit{i}} and the weighted variance $\sigma_i^2$ (Table~\ref{tab:gtr}) by
    \begin{equation}
        \begin{split}
            \vidx{d}{GTR\ \mathit{i}} &= \frac{\sum_{k=1}^5 (\sigma^{-2}_i)_k\, (\vidx{d}{GTR\ \mathit{i}})_k}{\sum_{k=1}^5 (\sigma^{-2}_i)_k}\quad , \\
            \sigma_i^2 &= \frac{1}{\sum_{k=1}^{5} (\sigma^2_i)_k}\quad .
        \end{split}
        \label{eq:dgtr}
    \end{equation}

    \begin{table}
        \caption{Weighted mean distance $\vidx{d}{GTR\ \mathit{i}}$ and weighted standard deviation $\vidx{\sigma}{GTR\ \mathit{i}}$ of the $i$-th GTR from the specular axis.}
        \label{tab:gtr}
        \begin{tabular}{cccc}
            \toprule
            \multicolumn{1}{c}{GTR} & \multicolumn{1}{c}{$\vidx{d}{GTR\ \mathit{i}}$} & \multicolumn{1}{c}{$\vidx{\sigma}{GTR\ \mathit{i}}$} & \multicolumn{1}{c}{$P$} \\
            \multicolumn{1}{c}{$i$} & \multicolumn{1}{c}{/ px} & \multicolumn{1}{c}{/ px} & \multicolumn{1}{c}{/ nm}\\ 
            \midrule
            $-3$ & 534.7 & 0.9 & 24.79\\
            $-2$ & 355.8 & 0.1 & 24.83\\
            $-1$ & 178.0 & 0.1 & 24.82\\
            $+1$ & 178.2 & 0.1 & 24.79\\
            $+2$ & 354.5 & 0.1 & 24.93\\
            $+3$ & 533.5 & 0.3 & 24.84\\
            \bottomrule
        \end{tabular}
    \end{table}

    \subsection{Traceable pitch determination}
    \label{sub:pitch}

    With the values and uncertainties for the photon energy \vidx{E}{ph} (section \ref{sub:instrum}), sample-detector distance \vidx{L}{s}, eq.~\eqref{eq:Ls}, pixel size \vidx{L}{px}, eq.~\eqref{eq:lpx}, and GTR distances (Table~\ref{tab:gtr}), the grating pitch $P$ can be calculated according to eq.~\eqref{eq:P2}.
    Note that the misalignment term $\cos\Delta\phi$ has been completely neglected as it deviates only by $10^{-10}$ from unity.
    The pitch $P$ is the average of the six pitches $P_i$ of the six GTR distances $(\vidx{d}{GTR})_i$.
    The combined standard uncertainty $u(P)$ is calculated according to the \emph{Guide to the Expression of Uncertainty in Measurement} (GUM) \cite{gum} from the uncertainty contributions of the constituting input parameters as shown in Table~\ref{tab:uncertainty}.
    In this way, the pitch of the self-assembled block copolymer grating is determined in a traceable way as
    \begin{equation}
        P = (24.83 \pm 0.09)~\mathrm{nm} .
        \label{eq:pitch-val}
    \end{equation}

    \begin{table*}
        \caption{GISAXS pitch uncertainty contributions $u(\vidx{x}{i})$ with their corresponding distribution (N = normal) and contribution type (A or B) according to GUM \cite{gum},
        uncertainty components $\vidx{u}{i}(\vidx{x}{i})$, relative uncertainty components $\frac{\vidx{u}{i}(\vidx{x}{i})}{\vidx{x}{i}}$
        and estimated combined standard uncertainty $\vidx{u}{c}(P)$ of the grating pitch $P = 24.83$~nm.}
        \label{tab:uncertainty}
        \begin{tabular}{lcrrc}
            \toprule
            \multicolumn{1}{c}{Input quantity \vidx{x}{i}} & \multicolumn{1}{c}{Distrib./type} & \multicolumn{1}{c}{$u(\vidx{x}{i})$}%
            & \multicolumn{1}{c}{$\frac{\vidx{u}{i}(\vidx{x}{i})}{\vidx{x}{i}}$} & \multicolumn{1}{c}{$\vidx{u}{c}(P)$ / nm} \\
            \midrule
            photon energy \vidx{E}{ph}              & N/B & 1~eV            & $1.0\times10^{-4}$    & 0.002 \\
            sample-detector distance \vidx{L}{s}    & N/B & 3~mm            & $1.0\times10^{-3}$    & 0.025\\
            pixel size \vidx{L}{px}                 & N/B & 0.2~\textmu m   & $2.9\times10^{-3}$    & 0.071\\
            GTR distances \vidx{d}{GTR}             & N/A & $\le 0.87$~px   & $\le1.6\times10^{-3}$ & 0.040\\
            \cmidrule(rl){1-5}
            \textbf{Combined standard uncertainty} $u(P)$ & & & & \textbf{0.09~nm}\\
            \bottomrule
        \end{tabular}
    \end{table*}

\section{Conclusions}
    In this work, the first traceable GISAXS determination of the pitch of a self-assembled block-copolymer line grating is presented.
    The deviation from the nominal value of only 25~nm is below 0.2~nm.
    Traceability is achieved by the uncertainty analysis of the contributing parameters: Sample-detector distance, detector pixel size, photon energy, and distance between the grating diffraction orders.
    The GISAXS patterns have been evaluated by forward data analysis of the GTR peak positions, which are exclusively defined by the grating structure factor.
    In this way, no additional assumptions on the form factor of the average line cross-section have to be made, which would not be traceable anymore as the correctness of the model itself cannot be evaluated from within the model.
    Hence, the basic experimental uncertainty of typical GISAXS measurements is determined with the presented kind of analysis.

    The predominant uncertainty contributions arise from the sample-detector distance and the pixel size, thus, the geometric parameters.
    Due to the high sensitivity of the GISAXS pattern to angular deviations $\Delta\phi$ from parallel alignment of incident beam and grating lines, the effect can be completely neglected in the uncertainty analysis once the grating is aligned.
    The analysis showed that the scattering spots of the GTRs have to be carefully examined for possible asymmetries, which were also found in the GISAXS images of the investigated sample.
    However, the imposed corrections were minor and two orders of magnitude below the combined standard uncertainty of the pitch in the present case.

    The presented uncertainty analysis may be used as an input parameter for more complex, but common GISAXS data analysis, for example within the framework of DWBA modeling.
    In this way, it might become possible to establish the traceability of structural parameters obtained from the numerical modeling of GISAXS data.
    
\acknowledgements
    The authors would like to thank Levent Cibik and Stefanie Langner (both from PTB) for their valuable assistance during the experiments, Melia Tjio and Joy Cheng (both from IBM Research) for providing the samples used in this work as well as Christian Gollwitzer (PTB) for the helpful discussions.

\bibliographystyle{aipnum4-1}
\bibliography{refs}

\begin{thebibliography}{62}%
\makeatletter
\providecommand \@ifxundefined [1]{%
 \@ifx{#1\undefined}
}%
\providecommand \@ifnum [1]{%
 \ifnum #1\expandafter \@firstoftwo
 \else \expandafter \@secondoftwo
 \fi
}%
\providecommand \@ifx [1]{%
 \ifx #1\expandafter \@firstoftwo
 \else \expandafter \@secondoftwo
 \fi
}%
\providecommand \natexlab [1]{#1}%
\providecommand \enquote  [1]{``#1''}%
\providecommand \bibnamefont  [1]{#1}%
\providecommand \bibfnamefont [1]{#1}%
\providecommand \citenamefont [1]{#1}%
\providecommand \href@noop [0]{\@secondoftwo}%
\providecommand \href [0]{\begingroup \@sanitize@url \@href}%
\providecommand \@href[1]{\@@startlink{#1}\@@href}%
\providecommand \@@href[1]{\endgroup#1\@@endlink}%
\providecommand \@sanitize@url [0]{\catcode `\\12\catcode `\$12\catcode
  `\&12\catcode `\#12\catcode `\^12\catcode `\_12\catcode `\%12\relax}%
\providecommand \@@startlink[1]{}%
\providecommand \@@endlink[0]{}%
\providecommand \url  [0]{\begingroup\@sanitize@url \@url }%
\providecommand \@url [1]{\endgroup\@href {#1}{\urlprefix }}%
\providecommand \urlprefix  [0]{URL }%
\providecommand \Eprint [0]{\href }%
\providecommand \doibase [0]{http://dx.doi.org/}%
\providecommand \selectlanguage [0]{\@gobble}%
\providecommand \bibinfo  [0]{\@secondoftwo}%
\providecommand \bibfield  [0]{\@secondoftwo}%
\providecommand \translation [1]{[#1]}%
\providecommand \BibitemOpen [0]{}%
\providecommand \bibitemStop [0]{}%
\providecommand \bibitemNoStop [0]{.\EOS\space}%
\providecommand \EOS [0]{\spacefactor3000\relax}%
\providecommand \BibitemShut  [1]{\csname bibitem#1\endcsname}%
\let\auto@bib@innerbib\@empty
\bibitem [{\citenamefont {Bhushan}(2010)}]{nanotech}%
  \BibitemOpen
  \bibfield  {author} {\bibinfo {author} {\bibfnamefont {B.}~\bibnamefont
  {Bhushan}},\ }\href@noop {} {\emph {\bibinfo {title} {{Springer Handbook of
  Nanotechnology}}}}\ (\bibinfo  {publisher} {Springer},\ \bibinfo {address}
  {Berlin, Heidelberg},\ \bibinfo {year} {2010})\BibitemShut {NoStop}%
\bibitem [{\citenamefont {Vogel}(2007)}]{vogel_tech_2007}%
  \BibitemOpen
  \bibfield  {author} {\bibinfo {author} {\bibfnamefont {E.}~\bibnamefont
  {Vogel}},\ }\href {\doibase 10.1038/nnano.2006.142} {\bibfield  {journal}
  {\bibinfo  {journal} {Nat. Nanotechnol.}\ }\textbf {\bibinfo {volume} {2}},\
  \bibinfo {pages} {25} (\bibinfo {year} {2007})}\BibitemShut {NoStop}%
\bibitem [{\citenamefont {Muller}(2005)}]{muller_sound_2005}%
  \BibitemOpen
  \bibfield  {author} {\bibinfo {author} {\bibfnamefont {D.}~\bibnamefont
  {Muller}},\ }\href {\doibase 10.1038/nmat1466} {\bibfield  {journal}
  {\bibinfo  {journal} {Nat. Mater.}\ }\textbf {\bibinfo {volume} {4}},\
  \bibinfo {pages} {645} (\bibinfo {year} {2005})}\BibitemShut {NoStop}%
\bibitem [{\citenamefont {Wagner}\ and\ \citenamefont
  {Harned}(2010)}]{wagner2010euv}%
  \BibitemOpen
  \bibfield  {author} {\bibinfo {author} {\bibfnamefont {C.}~\bibnamefont
  {Wagner}}\ and\ \bibinfo {author} {\bibfnamefont {N.}~\bibnamefont
  {Harned}},\ }\href@noop {} {\bibfield  {journal} {\bibinfo  {journal} {Nat.
  Photonics}\ }\textbf {\bibinfo {volume} {4}},\ \bibinfo {pages} {24}
  (\bibinfo {year} {2010})}\BibitemShut {NoStop}%
\bibitem [{\citenamefont {Leibler}(1980)}]{leibler1980bcptheory}%
  \BibitemOpen
  \bibfield  {author} {\bibinfo {author} {\bibfnamefont {L.}~\bibnamefont
  {Leibler}},\ }\href@noop {} {\bibfield  {journal} {\bibinfo  {journal}
  {Macromolecules}\ }\textbf {\bibinfo {volume} {13}},\ \bibinfo {pages} {1602}
  (\bibinfo {year} {1980})}\BibitemShut {NoStop}%
\bibitem [{\citenamefont {Darling}(2007)}]{darling2007directing}%
  \BibitemOpen
  \bibfield  {author} {\bibinfo {author} {\bibfnamefont {S.~B.}\ \bibnamefont
  {Darling}},\ }\href {\doibase 10.1016/j.progpolymsci.2007.05.004} {\bibfield
  {journal} {\bibinfo  {journal} {Progr. Polym. Sci.}\ }\textbf {\bibinfo
  {volume} {32}},\ \bibinfo {pages} {1152} (\bibinfo {year}
  {2007})}\BibitemShut {NoStop}%
\bibitem [{\citenamefont {Albert}\ and\ \citenamefont
  {Epps~{III}}(2010)}]{albert2010self}%
  \BibitemOpen
  \bibfield  {author} {\bibinfo {author} {\bibfnamefont {J.~N.~L.}\
  \bibnamefont {Albert}}\ and\ \bibinfo {author} {\bibfnamefont {T.~H.}\
  \bibnamefont {Epps~{III}}},\ }\href {\doibase 10.1016/S1369-7021(10)70106-1}
  {\bibfield  {journal} {\bibinfo  {journal} {Mater. Today}\ }\textbf {\bibinfo
  {volume} {13}},\ \bibinfo {pages} {24} (\bibinfo {year} {2010})}\BibitemShut
  {NoStop}%
\bibitem [{\citenamefont {Liu}\ \emph {et~al.}(2013)\citenamefont {Liu},
  \citenamefont {Ramirez-Hernandez}, \citenamefont {Han}, \citenamefont
  {Craig}, \citenamefont {Tada}, \citenamefont {Yoshida}, \citenamefont {Kang},
  \citenamefont {Ji}, \citenamefont {Gopalan}, \citenamefont {de~Pablo},\ and\
  \citenamefont {Nealey}}]{liu2013chemical}%
  \BibitemOpen
  \bibfield  {author} {\bibinfo {author} {\bibfnamefont {C.-C.}\ \bibnamefont
  {Liu}}, \bibinfo {author} {\bibfnamefont {A.}~\bibnamefont
  {Ramirez-Hernandez}}, \bibinfo {author} {\bibfnamefont {E.}~\bibnamefont
  {Han}}, \bibinfo {author} {\bibfnamefont {G.~S.~W.}\ \bibnamefont {Craig}},
  \bibinfo {author} {\bibfnamefont {Y.}~\bibnamefont {Tada}}, \bibinfo {author}
  {\bibfnamefont {H.}~\bibnamefont {Yoshida}}, \bibinfo {author} {\bibfnamefont
  {H.}~\bibnamefont {Kang}}, \bibinfo {author} {\bibfnamefont {S.}~\bibnamefont
  {Ji}}, \bibinfo {author} {\bibfnamefont {P.}~\bibnamefont {Gopalan}},
  \bibinfo {author} {\bibfnamefont {J.~J.}\ \bibnamefont {de~Pablo}}, \ and\
  \bibinfo {author} {\bibfnamefont {P.~F.}\ \bibnamefont {Nealey}},\ }\href
  {\doibase 10.1021/ma302464n} {\bibfield  {journal} {\bibinfo  {journal}
  {Macromolecules}\ }\textbf {\bibinfo {volume} {46}},\ \bibinfo {pages} {1415}
  (\bibinfo {year} {2013})}\BibitemShut {NoStop}%
\bibitem [{\citenamefont {Chang}\ \emph {et~al.}(2014)\citenamefont {Chang},
  \citenamefont {Choi}, \citenamefont {Hannon}, \citenamefont {Alexander-Katz},
  \citenamefont {Ross},\ and\ \citenamefont {Berggren}}]{chang_design_2014}%
  \BibitemOpen
  \bibfield  {author} {\bibinfo {author} {\bibfnamefont {J.-B.}\ \bibnamefont
  {Chang}}, \bibinfo {author} {\bibfnamefont {H.~K.}\ \bibnamefont {Choi}},
  \bibinfo {author} {\bibfnamefont {A.~F.}\ \bibnamefont {Hannon}}, \bibinfo
  {author} {\bibfnamefont {A.}~\bibnamefont {Alexander-Katz}}, \bibinfo
  {author} {\bibfnamefont {C.~A.}\ \bibnamefont {Ross}}, \ and\ \bibinfo
  {author} {\bibfnamefont {K.~K.}\ \bibnamefont {Berggren}},\ }\href {\doibase
  10.1038/ncomms4305} {\bibfield  {journal} {\bibinfo  {journal} {Nat.
  Commun.}\ }\textbf {\bibinfo {volume} {5}} (\bibinfo {year} {2014}),\
  10.1038/ncomms4305}\BibitemShut {NoStop}%
\bibitem [{\citenamefont {Bates}\ \emph {et~al.}(2014)\citenamefont {Bates},
  \citenamefont {Maher}, \citenamefont {Janes}, \citenamefont {Ellison},\ and\
  \citenamefont {Willson}}]{bates2014}%
  \BibitemOpen
  \bibfield  {author} {\bibinfo {author} {\bibfnamefont {C.}~\bibnamefont
  {Bates}}, \bibinfo {author} {\bibfnamefont {M.~J.}\ \bibnamefont {Maher}},
  \bibinfo {author} {\bibfnamefont {D.}~\bibnamefont {Janes}}, \bibinfo
  {author} {\bibfnamefont {C.}~\bibnamefont {Ellison}}, \ and\ \bibinfo
  {author} {\bibfnamefont {C.}~\bibnamefont {Willson}},\ }\href {\doibase
  10.1021/ma401762n} {\bibfield  {journal} {\bibinfo  {journal}
  {Macromolecules}\ }\textbf {\bibinfo {volume} {47}},\ \bibinfo {pages} {2}
  (\bibinfo {year} {2014})}\BibitemShut {NoStop}%
\bibitem [{\citenamefont {Hamley}(2003)}]{hamley2003nanofab}%
  \BibitemOpen
  \bibfield  {author} {\bibinfo {author} {\bibfnamefont {I.}~\bibnamefont
  {Hamley}},\ }\href@noop {} {\bibfield  {journal} {\bibinfo  {journal}
  {Nanotechnology}\ }\textbf {\bibinfo {volume} {14}},\ \bibinfo {pages} {R39}
  (\bibinfo {year} {2003})}\BibitemShut {NoStop}%
\bibitem [{\citenamefont {Saavedra}\ \emph {et~al.}(2010)\citenamefont
  {Saavedra}, \citenamefont {Mullen}, \citenamefont {Zhang}, \citenamefont
  {Dewey}, \citenamefont {Claridge},\ and\ \citenamefont
  {Weiss}}]{saavedra_hybrid_2010}%
  \BibitemOpen
  \bibfield  {author} {\bibinfo {author} {\bibfnamefont {H.~M.}\ \bibnamefont
  {Saavedra}}, \bibinfo {author} {\bibfnamefont {T.~J.}\ \bibnamefont
  {Mullen}}, \bibinfo {author} {\bibfnamefont {P.}~\bibnamefont {Zhang}},
  \bibinfo {author} {\bibfnamefont {D.~C.}\ \bibnamefont {Dewey}}, \bibinfo
  {author} {\bibfnamefont {S.~A.}\ \bibnamefont {Claridge}}, \ and\ \bibinfo
  {author} {\bibfnamefont {P.~S.}\ \bibnamefont {Weiss}},\ }\href {\doibase
  10.1088/0034-4885/73/3/036501} {\bibfield  {journal} {\bibinfo  {journal}
  {Rep. Progr. Phys.}\ }\textbf {\bibinfo {volume} {73}},\ \bibinfo {pages}
  {036501} (\bibinfo {year} {2010})}\BibitemShut {NoStop}%
\bibitem [{\citenamefont {Brabec}, \citenamefont {Scherf},\ and\ \citenamefont
  {Dyakonov}(2014)}]{brabec2014organic}%
  \BibitemOpen
  \bibfield  {author} {\bibinfo {author} {\bibfnamefont {C.}~\bibnamefont
  {Brabec}}, \bibinfo {author} {\bibfnamefont {U.}~\bibnamefont {Scherf}}, \
  and\ \bibinfo {author} {\bibfnamefont {V.}~\bibnamefont {Dyakonov}},\
  }\href@noop {} {\emph {\bibinfo {title} {Organic photovoltaics: materials,
  device physics, and manufacturing technologies}}}\ (\bibinfo  {publisher}
  {John Wiley \& Sons},\ \bibinfo {year} {2014})\BibitemShut {NoStop}%
\bibitem [{\citenamefont {Schaffer}\ \emph {et~al.}(2013)\citenamefont
  {Schaffer}, \citenamefont {Palumbiny}, \citenamefont {Niedermeier},
  \citenamefont {Jendrzejewski}, \citenamefont {Santoro}, \citenamefont
  {Roth},\ and\ \citenamefont {Müller-Buschbaum}}]{schaffer_direct_2013}%
  \BibitemOpen
  \bibfield  {author} {\bibinfo {author} {\bibfnamefont {C.~J.}\ \bibnamefont
  {Schaffer}}, \bibinfo {author} {\bibfnamefont {C.~M.}\ \bibnamefont
  {Palumbiny}}, \bibinfo {author} {\bibfnamefont {M.~A.}\ \bibnamefont
  {Niedermeier}}, \bibinfo {author} {\bibfnamefont {C.}~\bibnamefont
  {Jendrzejewski}}, \bibinfo {author} {\bibfnamefont {G.}~\bibnamefont
  {Santoro}}, \bibinfo {author} {\bibfnamefont {S.~V.}\ \bibnamefont {Roth}}, \
  and\ \bibinfo {author} {\bibfnamefont {P.}~\bibnamefont
  {Müller-Buschbaum}},\ }\href {\doibase 10.1002/adma.201302854} {\bibfield
  {journal} {\bibinfo  {journal} {Adv. Mater.}\ }\textbf {\bibinfo {volume}
  {25}},\ \bibinfo {pages} {6760–6764} (\bibinfo {year} {2013})}\BibitemShut
  {NoStop}%
\bibitem [{\citenamefont {Pechkova}\ \emph {et~al.}(2005)\citenamefont
  {Pechkova}, \citenamefont {Roth}, \citenamefont {Burghammer}, \citenamefont
  {Fontani}, \citenamefont {Riekel},\ and\ \citenamefont
  {Nicolini}}]{pechkova2005}%
  \BibitemOpen
  \bibfield  {author} {\bibinfo {author} {\bibfnamefont {E.}~\bibnamefont
  {Pechkova}}, \bibinfo {author} {\bibfnamefont {S.}~\bibnamefont {Roth}},
  \bibinfo {author} {\bibfnamefont {M.}~\bibnamefont {Burghammer}}, \bibinfo
  {author} {\bibfnamefont {D.}~\bibnamefont {Fontani}}, \bibinfo {author}
  {\bibfnamefont {C.}~\bibnamefont {Riekel}}, \ and\ \bibinfo {author}
  {\bibfnamefont {C.}~\bibnamefont {Nicolini}},\ }\href {\doibase
  10.1107/S0909049505011684} {\bibfield  {journal} {\bibinfo  {journal} {J.
  Synchrotron Radiat.}\ }\textbf {\bibinfo {volume} {12}},\ \bibinfo {pages}
  {713} (\bibinfo {year} {2005})}\BibitemShut {NoStop}%
\bibitem [{\citenamefont {Nie}\ and\ \citenamefont
  {Kumacheva}(2008)}]{nie_patterning_2008}%
  \BibitemOpen
  \bibfield  {author} {\bibinfo {author} {\bibfnamefont {Z.}~\bibnamefont
  {Nie}}\ and\ \bibinfo {author} {\bibfnamefont {E.}~\bibnamefont
  {Kumacheva}},\ }\href {\doibase 10.1038/nmat2109} {\bibfield  {journal}
  {\bibinfo  {journal} {Nat. Mater.}\ }\textbf {\bibinfo {volume} {7}},\
  \bibinfo {pages} {277} (\bibinfo {year} {2008})}\BibitemShut {NoStop}%
\bibitem [{\citenamefont {R{\"o}sler}, \citenamefont {Vandermeulen},\ and\
  \citenamefont {Klok}(2012)}]{rosler2012advanced}%
  \BibitemOpen
  \bibfield  {author} {\bibinfo {author} {\bibfnamefont {A.}~\bibnamefont
  {R{\"o}sler}}, \bibinfo {author} {\bibfnamefont {G.~W.}\ \bibnamefont
  {Vandermeulen}}, \ and\ \bibinfo {author} {\bibfnamefont {H.-A.}\
  \bibnamefont {Klok}},\ }\href@noop {} {\bibfield  {journal} {\bibinfo
  {journal} {Advanced drug delivery reviews}\ }\textbf {\bibinfo {volume}
  {64}},\ \bibinfo {pages} {270} (\bibinfo {year} {2012})}\BibitemShut
  {NoStop}%
\bibitem [{\citenamefont {Husemann}\ \emph {et~al.}(2000)\citenamefont
  {Husemann}, \citenamefont {Morrison}, \citenamefont {Benoit}, \citenamefont
  {Frommer}, \citenamefont {Mate}, \citenamefont {Hinsberg}, \citenamefont
  {Hedrick},\ and\ \citenamefont {Hawker}}]{husemann2000}%
  \BibitemOpen
  \bibfield  {author} {\bibinfo {author} {\bibfnamefont {M.}~\bibnamefont
  {Husemann}}, \bibinfo {author} {\bibfnamefont {M.}~\bibnamefont {Morrison}},
  \bibinfo {author} {\bibfnamefont {D.}~\bibnamefont {Benoit}}, \bibinfo
  {author} {\bibfnamefont {J.}~\bibnamefont {Frommer}}, \bibinfo {author}
  {\bibfnamefont {C.}~\bibnamefont {Mate}}, \bibinfo {author} {\bibfnamefont
  {W.}~\bibnamefont {Hinsberg}}, \bibinfo {author} {\bibfnamefont
  {J.}~\bibnamefont {Hedrick}}, \ and\ \bibinfo {author} {\bibfnamefont
  {C.}~\bibnamefont {Hawker}},\ }\href@noop {} {\bibfield  {journal} {\bibinfo
  {journal} {J. Am. Chem. Soc.}\ }\textbf {\bibinfo {volume} {122}},\ \bibinfo
  {pages} {1844} (\bibinfo {year} {2000})}\BibitemShut {NoStop}%
\bibitem [{\citenamefont {Leach}\ \emph {et~al.}(2011)\citenamefont {Leach},
  \citenamefont {Boyd}, \citenamefont {Burke}, \citenamefont {Danzebrink},
  \citenamefont {Dirscherl}, \citenamefont {Dziomba}, \citenamefont {Gee},
  \citenamefont {Koenders}, \citenamefont {Morazzani}, \citenamefont {Pidduck}
  \emph {et~al.}}]{leach2011european}%
  \BibitemOpen
  \bibfield  {author} {\bibinfo {author} {\bibfnamefont {R.~K.}\ \bibnamefont
  {Leach}}, \bibinfo {author} {\bibfnamefont {R.}~\bibnamefont {Boyd}},
  \bibinfo {author} {\bibfnamefont {T.}~\bibnamefont {Burke}}, \bibinfo
  {author} {\bibfnamefont {H.-U.}\ \bibnamefont {Danzebrink}}, \bibinfo
  {author} {\bibfnamefont {K.}~\bibnamefont {Dirscherl}}, \bibinfo {author}
  {\bibfnamefont {T.}~\bibnamefont {Dziomba}}, \bibinfo {author} {\bibfnamefont
  {M.}~\bibnamefont {Gee}}, \bibinfo {author} {\bibfnamefont {L.}~\bibnamefont
  {Koenders}}, \bibinfo {author} {\bibfnamefont {V.}~\bibnamefont {Morazzani}},
  \bibinfo {author} {\bibfnamefont {A.}~\bibnamefont {Pidduck}},  \emph
  {et~al.},\ }\href@noop {} {\bibfield  {journal} {\bibinfo  {journal}
  {Nanotechnology}\ }\textbf {\bibinfo {volume} {22}},\ \bibinfo {pages}
  {062001} (\bibinfo {year} {2011})}\BibitemShut {NoStop}%
\bibitem [{\citenamefont {Bosse}\ \emph {et~al.}(2009)\citenamefont {Bosse},
  \citenamefont {Boyd}, \citenamefont {Brand}, \citenamefont {Burke},
  \citenamefont {Cuenat}, \citenamefont {Danzebrink}, \citenamefont {Dircherl},
  \citenamefont {Dziomba}, \citenamefont {Fl{\"u}gge}, \citenamefont {Frase}
  \emph {et~al.}}]{co-nanomet-foresight}%
  \BibitemOpen
  \bibfield  {author} {\bibinfo {author} {\bibfnamefont {H.}~\bibnamefont
  {Bosse}}, \bibinfo {author} {\bibfnamefont {R.}~\bibnamefont {Boyd}},
  \bibinfo {author} {\bibfnamefont {U.}~\bibnamefont {Brand}}, \bibinfo
  {author} {\bibfnamefont {T.}~\bibnamefont {Burke}}, \bibinfo {author}
  {\bibfnamefont {A.}~\bibnamefont {Cuenat}}, \bibinfo {author} {\bibfnamefont
  {H.}~\bibnamefont {Danzebrink}}, \bibinfo {author} {\bibfnamefont
  {K.}~\bibnamefont {Dircherl}}, \bibinfo {author} {\bibfnamefont
  {T.}~\bibnamefont {Dziomba}}, \bibinfo {author} {\bibfnamefont
  {J.}~\bibnamefont {Fl{\"u}gge}}, \bibinfo {author} {\bibfnamefont
  {G.}~\bibnamefont {Frase}},  \emph {et~al.},\ }\href@noop {} {\enquote
  {\bibinfo {title} {Nanometrology foresight review (co-nanomet project output
  www.co-nanomet.eu)},}\ } (\bibinfo {year} {2009})\BibitemShut {NoStop}%
\bibitem [{\citenamefont {Bunday}\ \emph {et~al.}(2003)\citenamefont {Bunday},
  \citenamefont {Bishop}, \citenamefont {Villarrubia},\ and\ \citenamefont
  {Vladar}}]{bunday2003cd}%
  \BibitemOpen
  \bibfield  {author} {\bibinfo {author} {\bibfnamefont {B.~D.}\ \bibnamefont
  {Bunday}}, \bibinfo {author} {\bibfnamefont {M.}~\bibnamefont {Bishop}},
  \bibinfo {author} {\bibfnamefont {J.~S.}\ \bibnamefont {Villarrubia}}, \ and\
  \bibinfo {author} {\bibfnamefont {A.~E.}\ \bibnamefont {Vladar}},\ }\href
  {\doibase 10.1117/12.485007} {\bibfield  {journal} {\bibinfo  {journal}
  {Proc. SPIE}\ }\textbf {\bibinfo {volume} {5038}},\ \bibinfo {pages} {674}
  (\bibinfo {year} {2003})}\BibitemShut {NoStop}%
\bibitem [{\citenamefont {Villarrubia}, \citenamefont {Vlad{\'a}r},\ and\
  \citenamefont {Postek}(2005)}]{villarrubia2005scanning}%
  \BibitemOpen
  \bibfield  {author} {\bibinfo {author} {\bibfnamefont {J.}~\bibnamefont
  {Villarrubia}}, \bibinfo {author} {\bibfnamefont {A.}~\bibnamefont
  {Vlad{\'a}r}}, \ and\ \bibinfo {author} {\bibfnamefont {M.}~\bibnamefont
  {Postek}},\ }\href@noop {} {\bibfield  {journal} {\bibinfo  {journal} {Surf.
  Interf. Anal.}\ }\textbf {\bibinfo {volume} {37}},\ \bibinfo {pages} {951}
  (\bibinfo {year} {2005})}\BibitemShut {NoStop}%
\bibitem [{\citenamefont {Misumi}\ \emph {et~al.}(2003)\citenamefont {Misumi},
  \citenamefont {Gonda}, \citenamefont {Kurosawa},\ and\ \citenamefont
  {Takamasu}}]{misumi2003uncertainty}%
  \BibitemOpen
  \bibfield  {author} {\bibinfo {author} {\bibfnamefont {I.}~\bibnamefont
  {Misumi}}, \bibinfo {author} {\bibfnamefont {S.}~\bibnamefont {Gonda}},
  \bibinfo {author} {\bibfnamefont {T.}~\bibnamefont {Kurosawa}}, \ and\
  \bibinfo {author} {\bibfnamefont {K.}~\bibnamefont {Takamasu}},\ }\href@noop
  {} {\bibfield  {journal} {\bibinfo  {journal} {Meas. Sci. Technol.}\ }\textbf
  {\bibinfo {volume} {14}},\ \bibinfo {pages} {463} (\bibinfo {year}
  {2003})}\BibitemShut {NoStop}%
\bibitem [{\citenamefont {Yacoot}\ and\ \citenamefont
  {Koenders}(2011)}]{yacoot2011recent}%
  \BibitemOpen
  \bibfield  {author} {\bibinfo {author} {\bibfnamefont {A.}~\bibnamefont
  {Yacoot}}\ and\ \bibinfo {author} {\bibfnamefont {L.}~\bibnamefont
  {Koenders}},\ }\href@noop {} {\bibfield  {journal} {\bibinfo  {journal}
  {Meas. Sci. Technol.}\ }\textbf {\bibinfo {volume} {22}},\ \bibinfo {pages}
  {122001} (\bibinfo {year} {2011})}\BibitemShut {NoStop}%
\bibitem [{\citenamefont {Krumrey}\ \emph {et~al.}(2011)\citenamefont
  {Krumrey}, \citenamefont {Gleber}, \citenamefont {Scholze},\ and\
  \citenamefont {Wernecke}}]{krumrey2011synchrotron}%
  \BibitemOpen
  \bibfield  {author} {\bibinfo {author} {\bibfnamefont {M.}~\bibnamefont
  {Krumrey}}, \bibinfo {author} {\bibfnamefont {G.}~\bibnamefont {Gleber}},
  \bibinfo {author} {\bibfnamefont {F.}~\bibnamefont {Scholze}}, \ and\
  \bibinfo {author} {\bibfnamefont {J.}~\bibnamefont {Wernecke}},\ }\href@noop
  {} {\bibfield  {journal} {\bibinfo  {journal} {Meas. Sci. Technol.}\ }\textbf
  {\bibinfo {volume} {22}},\ \bibinfo {pages} {094032} (\bibinfo {year}
  {2011})}\BibitemShut {NoStop}%
\bibitem [{\citenamefont {Wernecke}, \citenamefont {Scholze},\ and\
  \citenamefont {Krumrey}(2012)}]{wernecke2012}%
  \BibitemOpen
  \bibfield  {author} {\bibinfo {author} {\bibfnamefont {J.}~\bibnamefont
  {Wernecke}}, \bibinfo {author} {\bibfnamefont {F.}~\bibnamefont {Scholze}}, \
  and\ \bibinfo {author} {\bibfnamefont {M.}~\bibnamefont {Krumrey}},\ }\href
  {\doibase 10.1063/1.4758283} {\bibfield  {journal} {\bibinfo  {journal} {Rev.
  Sci. Instrum.}\ }\textbf {\bibinfo {volume} {83}},\ \bibinfo {pages} {103906}
  (\bibinfo {year} {2012})}\BibitemShut {NoStop}%
\bibitem [{\citenamefont {Hofmann}, \citenamefont {Dobisz},\ and\ \citenamefont
  {Ocko}(2009)}]{hofmann2009}%
  \BibitemOpen
  \bibfield  {author} {\bibinfo {author} {\bibfnamefont {T.}~\bibnamefont
  {Hofmann}}, \bibinfo {author} {\bibfnamefont {E.}~\bibnamefont {Dobisz}}, \
  and\ \bibinfo {author} {\bibfnamefont {B.~M.}\ \bibnamefont {Ocko}},\ }\href
  {\doibase 10.1116/1.3253608} {\bibfield  {journal} {\bibinfo  {journal} {J.
  Vac. Sci. Technol. B}\ }\textbf {\bibinfo {volume} {27}},\ \bibinfo {pages}
  {3238} (\bibinfo {year} {2009})}\BibitemShut {NoStop}%
\bibitem [{\citenamefont {Jones}\ \emph {et~al.}(2003)\citenamefont {Jones},
  \citenamefont {Hu}, \citenamefont {Lin}, \citenamefont {Wu}, \citenamefont
  {Kolb}, \citenamefont {Casa}, \citenamefont {Bolton},\ and\ \citenamefont
  {Barclay}}]{jones2003}%
  \BibitemOpen
  \bibfield  {author} {\bibinfo {author} {\bibfnamefont {R.~L.}\ \bibnamefont
  {Jones}}, \bibinfo {author} {\bibfnamefont {T.}~\bibnamefont {Hu}}, \bibinfo
  {author} {\bibfnamefont {E.~K.}\ \bibnamefont {Lin}}, \bibinfo {author}
  {\bibfnamefont {W.-l.}\ \bibnamefont {Wu}}, \bibinfo {author} {\bibfnamefont
  {R.}~\bibnamefont {Kolb}}, \bibinfo {author} {\bibfnamefont {D.~M.}\
  \bibnamefont {Casa}}, \bibinfo {author} {\bibfnamefont {P.~J.}\ \bibnamefont
  {Bolton}}, \ and\ \bibinfo {author} {\bibfnamefont {G.~G.}\ \bibnamefont
  {Barclay}},\ }\href {\doibase http://dx.doi.org/10.1063/1.1622793} {\bibfield
   {journal} {\bibinfo  {journal} {Appl. Phys. Lett.}\ }\textbf {\bibinfo
  {volume} {83}},\ \bibinfo {pages} {4059} (\bibinfo {year}
  {2003})}\BibitemShut {NoStop}%
\bibitem [{\citenamefont {Hu}\ \emph {et~al.}(2004)\citenamefont {Hu},
  \citenamefont {Jones}, \citenamefont {Wu}, \citenamefont {Lin}, \citenamefont
  {Lin}, \citenamefont {Keane}, \citenamefont {Weigand},\ and\ \citenamefont
  {Quintana}}]{hu2004}%
  \BibitemOpen
  \bibfield  {author} {\bibinfo {author} {\bibfnamefont {T.}~\bibnamefont
  {Hu}}, \bibinfo {author} {\bibfnamefont {R.~L.}\ \bibnamefont {Jones}},
  \bibinfo {author} {\bibfnamefont {W.-l.}\ \bibnamefont {Wu}}, \bibinfo
  {author} {\bibfnamefont {E.~K.}\ \bibnamefont {Lin}}, \bibinfo {author}
  {\bibfnamefont {Q.}~\bibnamefont {Lin}}, \bibinfo {author} {\bibfnamefont
  {D.}~\bibnamefont {Keane}}, \bibinfo {author} {\bibfnamefont
  {S.}~\bibnamefont {Weigand}}, \ and\ \bibinfo {author} {\bibfnamefont
  {J.}~\bibnamefont {Quintana}},\ }\href {\doibase
  http://dx.doi.org/10.1063/1.1773376} {\bibfield  {journal} {\bibinfo
  {journal} {J. Appl. Phys.}\ }\textbf {\bibinfo {volume} {96}},\ \bibinfo
  {pages} {1983} (\bibinfo {year} {2004})}\BibitemShut {NoStop}%
\bibitem [{\citenamefont {Sunday}\ \emph {et~al.}(ress)\citenamefont {Sunday},
  \citenamefont {Hammond}, \citenamefont {Wang}, \citenamefont {Wu},
  \citenamefont {Delongchamp}, \citenamefont {Tijo}, \citenamefont {Cheng},
  \citenamefont {Pitera},\ and\ \citenamefont {Kline}}]{Sunday2014}%
  \BibitemOpen
  \bibfield  {author} {\bibinfo {author} {\bibfnamefont {D.~F.}\ \bibnamefont
  {Sunday}}, \bibinfo {author} {\bibfnamefont {M.~R.}\ \bibnamefont {Hammond}},
  \bibinfo {author} {\bibfnamefont {C.}~\bibnamefont {Wang}}, \bibinfo {author}
  {\bibfnamefont {W.-l.}\ \bibnamefont {Wu}}, \bibinfo {author} {\bibfnamefont
  {D.~M.}\ \bibnamefont {Delongchamp}}, \bibinfo {author} {\bibfnamefont
  {M.}~\bibnamefont {Tijo}}, \bibinfo {author} {\bibfnamefont {J.~Y.}\
  \bibnamefont {Cheng}}, \bibinfo {author} {\bibfnamefont {J.~W.}\ \bibnamefont
  {Pitera}}, \ and\ \bibinfo {author} {\bibfnamefont {R.~J.}\ \bibnamefont
  {Kline}},\ }\href@noop {} {\bibfield  {journal} {\bibinfo  {journal} {ACS
  Nano}\ } (\bibinfo {year} {in press})}\BibitemShut {NoStop}%
\bibitem [{\citenamefont {Gross}\ \emph {et~al.}(2009)\citenamefont {Gross},
  \citenamefont {Rathsfeld}, \citenamefont {Scholze},\ and\ \citenamefont
  {B{\"a}r}}]{gross2009profile}%
  \BibitemOpen
  \bibfield  {author} {\bibinfo {author} {\bibfnamefont {H.}~\bibnamefont
  {Gross}}, \bibinfo {author} {\bibfnamefont {A.}~\bibnamefont {Rathsfeld}},
  \bibinfo {author} {\bibfnamefont {F.}~\bibnamefont {Scholze}}, \ and\
  \bibinfo {author} {\bibfnamefont {M.}~\bibnamefont {B{\"a}r}},\ }\href@noop
  {} {\bibfield  {journal} {\bibinfo  {journal} {Meas. Sci. Technol.}\ }\textbf
  {\bibinfo {volume} {20}},\ \bibinfo {pages} {105102} (\bibinfo {year}
  {2009})}\BibitemShut {NoStop}%
\bibitem [{\citenamefont {Perlich}\ \emph {et~al.}(2004)\citenamefont
  {Perlich}, \citenamefont {Kamm}, \citenamefont {Rau}, \citenamefont
  {Scholze},\ and\ \citenamefont {Ulm}}]{perlich2004characterization}%
  \BibitemOpen
  \bibfield  {author} {\bibinfo {author} {\bibfnamefont {J.}~\bibnamefont
  {Perlich}}, \bibinfo {author} {\bibfnamefont {F.-M.}\ \bibnamefont {Kamm}},
  \bibinfo {author} {\bibfnamefont {J.}~\bibnamefont {Rau}}, \bibinfo {author}
  {\bibfnamefont {F.}~\bibnamefont {Scholze}}, \ and\ \bibinfo {author}
  {\bibfnamefont {G.}~\bibnamefont {Ulm}},\ }\href@noop {} {\bibfield
  {journal} {\bibinfo  {journal} {J. Vac. Sci. Technol. B}\ }\textbf {\bibinfo
  {volume} {22}},\ \bibinfo {pages} {3059} (\bibinfo {year}
  {2004})}\BibitemShut {NoStop}%
\bibitem [{\citenamefont {Levine}\ \emph {et~al.}(1989)\citenamefont {Levine},
  \citenamefont {Cohen}, \citenamefont {Chung},\ and\ \citenamefont
  {Georgopoulos}}]{levine1989}%
  \BibitemOpen
  \bibfield  {author} {\bibinfo {author} {\bibfnamefont {J.~R.}\ \bibnamefont
  {Levine}}, \bibinfo {author} {\bibfnamefont {J.~B.}\ \bibnamefont {Cohen}},
  \bibinfo {author} {\bibfnamefont {Y.~W.}\ \bibnamefont {Chung}}, \ and\
  \bibinfo {author} {\bibfnamefont {P.}~\bibnamefont {Georgopoulos}},\ }\href
  {\doibase 10.1107/S002188988900717X} {\bibfield  {journal} {\bibinfo
  {journal} {J. Appl. Cryst.}\ }\textbf {\bibinfo {volume} {22}},\ \bibinfo
  {pages} {528} (\bibinfo {year} {1989})}\BibitemShut {NoStop}%
\bibitem [{\citenamefont {Renaud}, \citenamefont {Lazzari},\ and\ \citenamefont
  {Leroy}(2009)}]{renaud2009}%
  \BibitemOpen
  \bibfield  {author} {\bibinfo {author} {\bibfnamefont {G.}~\bibnamefont
  {Renaud}}, \bibinfo {author} {\bibfnamefont {R.}~\bibnamefont {Lazzari}}, \
  and\ \bibinfo {author} {\bibfnamefont {F.}~\bibnamefont {Leroy}},\
  }\href@noop {} {\bibfield  {journal} {\bibinfo  {journal} {Surf. Sci. Rep.}\
  }\textbf {\bibinfo {volume} {64}},\ \bibinfo {pages} {255} (\bibinfo {year}
  {2009})}\BibitemShut {NoStop}%
\bibitem [{\citenamefont {M\"{u}ller-Buschbaum}(2003)}]{pmb2003}%
  \BibitemOpen
  \bibfield  {author} {\bibinfo {author} {\bibfnamefont {P.}~\bibnamefont
  {M\"{u}ller-Buschbaum}},\ }\href@noop {} {\bibfield  {journal} {\bibinfo
  {journal} {Anal. Bioanal. Chem.}\ }\textbf {\bibinfo {volume} {376}},\
  \bibinfo {pages} {3} (\bibinfo {year} {2003})}\BibitemShut {NoStop}%
\bibitem [{\citenamefont {Lee}\ \emph {et~al.}(2005)\citenamefont {Lee},
  \citenamefont {Park}, \citenamefont {Yoon}, \citenamefont {Park},
  \citenamefont {Kim}, \citenamefont {Kim}, \citenamefont {Chang},\ and\
  \citenamefont {Ree}}]{lee2005}%
  \BibitemOpen
  \bibfield  {author} {\bibinfo {author} {\bibfnamefont {B.}~\bibnamefont
  {Lee}}, \bibinfo {author} {\bibfnamefont {I.}~\bibnamefont {Park}}, \bibinfo
  {author} {\bibfnamefont {J.}~\bibnamefont {Yoon}}, \bibinfo {author}
  {\bibfnamefont {S.}~\bibnamefont {Park}}, \bibinfo {author} {\bibfnamefont
  {J.}~\bibnamefont {Kim}}, \bibinfo {author} {\bibfnamefont {K.}~\bibnamefont
  {Kim}}, \bibinfo {author} {\bibfnamefont {T.}~\bibnamefont {Chang}}, \ and\
  \bibinfo {author} {\bibfnamefont {M.}~\bibnamefont {Ree}},\ }\href@noop {}
  {\bibfield  {journal} {\bibinfo  {journal} {Macromolecules}\ }\textbf
  {\bibinfo {volume} {38}},\ \bibinfo {pages} {4311} (\bibinfo {year}
  {2005})}\BibitemShut {NoStop}%
\bibitem [{\citenamefont {M\"{u}ller-Buschbaum}(2008)}]{stamm2008}%
  \BibitemOpen
  \bibfield  {author} {\bibinfo {author} {\bibfnamefont {P.}~\bibnamefont
  {M\"{u}ller-Buschbaum}},\ }\href@noop {} {\emph {\bibinfo {title} {Structure
  Determination in Thin Film Geometry Using Grazing Incidence Small-Angle
  Scattering; In: {Polymer Surfaces and Interfaces}}}},\ edited by\ \bibinfo
  {editor} {\bibfnamefont {M.}~\bibnamefont {Stamm}}\ (\bibinfo  {publisher}
  {Springer},\ \bibinfo {address} {Berlin, Heidelberg},\ \bibinfo {year}
  {2008})\ pp.\ \bibinfo {pages} {17--46}\BibitemShut {NoStop}%
\bibitem [{\citenamefont {Wang}\ \emph {et~al.}(2011)\citenamefont {Wang},
  \citenamefont {Lee}, \citenamefont {Hexemer}, \citenamefont {Kim},
  \citenamefont {Zhao}, \citenamefont {Hasegawa}, \citenamefont {Ade},\ and\
  \citenamefont {Russell}}]{wang2011}%
  \BibitemOpen
  \bibfield  {author} {\bibinfo {author} {\bibfnamefont {C.}~\bibnamefont
  {Wang}}, \bibinfo {author} {\bibfnamefont {D.}~\bibnamefont {Lee}}, \bibinfo
  {author} {\bibfnamefont {A.}~\bibnamefont {Hexemer}}, \bibinfo {author}
  {\bibfnamefont {M.}~\bibnamefont {Kim}}, \bibinfo {author} {\bibfnamefont
  {W.}~\bibnamefont {Zhao}}, \bibinfo {author} {\bibfnamefont {H.}~\bibnamefont
  {Hasegawa}}, \bibinfo {author} {\bibfnamefont {H.}~\bibnamefont {Ade}}, \
  and\ \bibinfo {author} {\bibfnamefont {T.}~\bibnamefont {Russell}},\
  }\href@noop {} {\bibfield  {journal} {\bibinfo  {journal} {Nano Lett.}\
  }\textbf {\bibinfo {volume} {11}},\ \bibinfo {pages} {3906} (\bibinfo {year}
  {2011})}\BibitemShut {NoStop}%
\bibitem [{\citenamefont {Okuda}\ \emph {et~al.}(2011)\citenamefont {Okuda},
  \citenamefont {Takeshita}, \citenamefont {Ochiai}, \citenamefont {Sakurai},\
  and\ \citenamefont {Kitajima}}]{okuda2011near}%
  \BibitemOpen
  \bibfield  {author} {\bibinfo {author} {\bibfnamefont {H.}~\bibnamefont
  {Okuda}}, \bibinfo {author} {\bibfnamefont {K.}~\bibnamefont {Takeshita}},
  \bibinfo {author} {\bibfnamefont {S.}~\bibnamefont {Ochiai}}, \bibinfo
  {author} {\bibfnamefont {S.-I.}\ \bibnamefont {Sakurai}}, \ and\ \bibinfo
  {author} {\bibfnamefont {Y.}~\bibnamefont {Kitajima}},\ }\href@noop {}
  {\bibfield  {journal} {\bibinfo  {journal} {J. Appl. Cryst.}\ }\textbf
  {\bibinfo {volume} {44}},\ \bibinfo {pages} {380} (\bibinfo {year}
  {2011})}\BibitemShut {NoStop}%
\bibitem [{\citenamefont {Hoydalsvik}\ \emph {et~al.}(2010)\citenamefont
  {Hoydalsvik}, \citenamefont {Barnardo}, \citenamefont {Winter}, \citenamefont
  {Haas}, \citenamefont {Tatchev},\ and\ \citenamefont
  {Hoell}}]{hoydalsvik2010}%
  \BibitemOpen
  \bibfield  {author} {\bibinfo {author} {\bibfnamefont {K.}~\bibnamefont
  {Hoydalsvik}}, \bibinfo {author} {\bibfnamefont {T.}~\bibnamefont
  {Barnardo}}, \bibinfo {author} {\bibfnamefont {R.}~\bibnamefont {Winter}},
  \bibinfo {author} {\bibfnamefont {S.}~\bibnamefont {Haas}}, \bibinfo {author}
  {\bibfnamefont {D.}~\bibnamefont {Tatchev}}, \ and\ \bibinfo {author}
  {\bibfnamefont {A.}~\bibnamefont {Hoell}},\ }\href@noop {} {\bibfield
  {journal} {\bibinfo  {journal} {Phys. Chem. Chem. Phys.}\ }\textbf {\bibinfo
  {volume} {12}},\ \bibinfo {pages} {14492} (\bibinfo {year}
  {2010})}\BibitemShut {NoStop}%
\bibitem [{\citenamefont {Son}\ \emph {et~al.}(2013)\citenamefont {Son},
  \citenamefont {Son}, \citenamefont {Moon}, \citenamefont {Lee}, \citenamefont
  {Myoung}, \citenamefont {Strano}, \citenamefont {Ham},\ and\ \citenamefont
  {Ross}}]{son2013sub}%
  \BibitemOpen
  \bibfield  {author} {\bibinfo {author} {\bibfnamefont {J.~G.}\ \bibnamefont
  {Son}}, \bibinfo {author} {\bibfnamefont {M.}~\bibnamefont {Son}}, \bibinfo
  {author} {\bibfnamefont {K.-J.}\ \bibnamefont {Moon}}, \bibinfo {author}
  {\bibfnamefont {B.~H.}\ \bibnamefont {Lee}}, \bibinfo {author} {\bibfnamefont
  {J.-M.}\ \bibnamefont {Myoung}}, \bibinfo {author} {\bibfnamefont {M.~S.}\
  \bibnamefont {Strano}}, \bibinfo {author} {\bibfnamefont {M.-H.}\
  \bibnamefont {Ham}}, \ and\ \bibinfo {author} {\bibfnamefont {C.~A.}\
  \bibnamefont {Ross}},\ }\href@noop {} {\bibfield  {journal} {\bibinfo
  {journal} {Adv. Mater.}\ }\textbf {\bibinfo {volume} {25}},\ \bibinfo {pages}
  {4723} (\bibinfo {year} {2013})}\BibitemShut {NoStop}%
\bibitem [{\citenamefont {Sinha}\ \emph {et~al.}(1988)\citenamefont {Sinha},
  \citenamefont {Sirota}, \citenamefont {Garoff},\ and\ \citenamefont
  {Stanley}}]{sinha1988}%
  \BibitemOpen
  \bibfield  {author} {\bibinfo {author} {\bibfnamefont {S.~K.}\ \bibnamefont
  {Sinha}}, \bibinfo {author} {\bibfnamefont {E.~B.}\ \bibnamefont {Sirota}},
  \bibinfo {author} {\bibfnamefont {S.}~\bibnamefont {Garoff}}, \ and\ \bibinfo
  {author} {\bibfnamefont {H.~B.}\ \bibnamefont {Stanley}},\ }\href {\doibase
  10.1103/PhysRevB.38.2297} {\bibfield  {journal} {\bibinfo  {journal} {Phys.
  Rev. B}\ }\textbf {\bibinfo {volume} {38}},\ \bibinfo {pages} {2297}
  (\bibinfo {year} {1988})}\BibitemShut {NoStop}%
\bibitem [{\citenamefont {Hol{\'y}}\ and\ \citenamefont
  {Baumbach}(1994)}]{holy1994}%
  \BibitemOpen
  \bibfield  {author} {\bibinfo {author} {\bibfnamefont {V.}~\bibnamefont
  {Hol{\'y}}}\ and\ \bibinfo {author} {\bibfnamefont {T.}~\bibnamefont
  {Baumbach}},\ }\href {\doibase 10.1103/PhysRevB.49.10668} {\bibfield
  {journal} {\bibinfo  {journal} {Phys. Rev. B}\ }\textbf {\bibinfo {volume}
  {49}},\ \bibinfo {pages} {10668} (\bibinfo {year} {1994})}\BibitemShut
  {NoStop}%
\bibitem [{\citenamefont {Salditt}\ \emph {et~al.}(1995)\citenamefont
  {Salditt}, \citenamefont {Metzger}, \citenamefont {Brandt}, \citenamefont
  {Klemradt},\ and\ \citenamefont {Peisl}}]{salditt1995}%
  \BibitemOpen
  \bibfield  {author} {\bibinfo {author} {\bibfnamefont {T.}~\bibnamefont
  {Salditt}}, \bibinfo {author} {\bibfnamefont {T.~H.}\ \bibnamefont
  {Metzger}}, \bibinfo {author} {\bibfnamefont {C.}~\bibnamefont {Brandt}},
  \bibinfo {author} {\bibfnamefont {U.}~\bibnamefont {Klemradt}}, \ and\
  \bibinfo {author} {\bibfnamefont {J.}~\bibnamefont {Peisl}},\ }\href
  {\doibase 10.1103/PhysRevB.51.5617} {\bibfield  {journal} {\bibinfo
  {journal} {Phys. Rev. B}\ }\textbf {\bibinfo {volume} {51}},\ \bibinfo
  {pages} {5617} (\bibinfo {year} {1995})}\BibitemShut {NoStop}%
\bibitem [{\citenamefont {Wu}(1993)}]{wu1993}%
  \BibitemOpen
  \bibfield  {author} {\bibinfo {author} {\bibfnamefont {W.-l.}\ \bibnamefont
  {Wu}},\ }\href {\doibase 10.1063/1.464284} {\bibfield  {journal} {\bibinfo
  {journal} {J. Chem. Phys.}\ }\textbf {\bibinfo {volume} {98}},\ \bibinfo
  {pages} {1687} (\bibinfo {year} {1993})}\BibitemShut {NoStop}%
\bibitem [{\citenamefont {Wu}(1994)}]{wu1994}%
  \BibitemOpen
  \bibfield  {author} {\bibinfo {author} {\bibfnamefont {W.-l.}\ \bibnamefont
  {Wu}},\ }\href {\doibase 10.1063/1.468464} {\bibfield  {journal} {\bibinfo
  {journal} {J. Chem. Phys.}\ }\textbf {\bibinfo {volume} {101}},\ \bibinfo
  {pages} {4198} (\bibinfo {year} {1994})}\BibitemShut {NoStop}%
\bibitem [{\citenamefont {Lazzari}(2002)}]{isgisaxs}%
  \BibitemOpen
  \bibfield  {author} {\bibinfo {author} {\bibfnamefont {R.}~\bibnamefont
  {Lazzari}},\ }\href {\doibase 10.1107/S0021889802006088} {\bibfield
  {journal} {\bibinfo  {journal} {J. Appl. Cryst.}\ }\textbf {\bibinfo {volume}
  {35}},\ \bibinfo {pages} {406} (\bibinfo {year} {2002})}\BibitemShut
  {NoStop}%
\bibitem [{\citenamefont {Babonneau}(2010)}]{fitgisaxs}%
  \BibitemOpen
  \bibfield  {author} {\bibinfo {author} {\bibfnamefont {D.}~\bibnamefont
  {Babonneau}},\ }\href {\doibase 10.1107/S0021889810020352} {\bibfield
  {journal} {\bibinfo  {journal} {J. Appl. Cryst.}\ }\textbf {\bibinfo {volume}
  {43}},\ \bibinfo {pages} {929} (\bibinfo {year} {2010})}\BibitemShut
  {NoStop}%
\bibitem [{\citenamefont {Jergel}\ \emph {et~al.}(1999)\citenamefont {Jergel},
  \citenamefont {Mikul{\'\i}k}, \citenamefont {Majkov{\'a}}, \citenamefont
  {Luby}, \citenamefont {Sender{\'a}k}, \citenamefont {Pinc{\'\i}k},
  \citenamefont {Brunel}, \citenamefont {Hudek}, \citenamefont {Kostic},\ and\
  \citenamefont {Konecn{\'\i}kov{\'a}}}]{jergel1999}%
  \BibitemOpen
  \bibfield  {author} {\bibinfo {author} {\bibfnamefont {M.}~\bibnamefont
  {Jergel}}, \bibinfo {author} {\bibfnamefont {P.}~\bibnamefont
  {Mikul{\'\i}k}}, \bibinfo {author} {\bibfnamefont {E.}~\bibnamefont
  {Majkov{\'a}}}, \bibinfo {author} {\bibfnamefont {S.}~\bibnamefont {Luby}},
  \bibinfo {author} {\bibfnamefont {R.}~\bibnamefont {Sender{\'a}k}}, \bibinfo
  {author} {\bibfnamefont {E.}~\bibnamefont {Pinc{\'\i}k}}, \bibinfo {author}
  {\bibfnamefont {M.}~\bibnamefont {Brunel}}, \bibinfo {author} {\bibfnamefont
  {P.}~\bibnamefont {Hudek}}, \bibinfo {author} {\bibfnamefont
  {I.}~\bibnamefont {Kostic}}, \ and\ \bibinfo {author} {\bibfnamefont
  {A.}~\bibnamefont {Konecn{\'\i}kov{\'a}}},\ }\href {\doibase
  10.1088/0022-3727/32/10A/343} {\bibfield  {journal} {\bibinfo  {journal} {J.
  Phys. D App. Phys.}\ }\textbf {\bibinfo {volume} {32}},\ \bibinfo {pages}
  {A220} (\bibinfo {year} {1999})}\BibitemShut {NoStop}%
\bibitem [{\citenamefont {Mikul{\'\i}k}\ \emph {et~al.}(2001)\citenamefont
  {Mikul{\'\i}k}, \citenamefont {Jergel}, \citenamefont {Baumbach},
  \citenamefont {Majkov{\'a}}, \citenamefont {Pinc{\'\i}k}, \citenamefont
  {Luby}, \citenamefont {Ortega}, \citenamefont {Tucoulou}, \citenamefont
  {Hudek},\ and\ \citenamefont {Kostic}}]{mikulik2001}%
  \BibitemOpen
  \bibfield  {author} {\bibinfo {author} {\bibfnamefont {P.}~\bibnamefont
  {Mikul{\'\i}k}}, \bibinfo {author} {\bibfnamefont {M.}~\bibnamefont
  {Jergel}}, \bibinfo {author} {\bibfnamefont {T.}~\bibnamefont {Baumbach}},
  \bibinfo {author} {\bibfnamefont {E.}~\bibnamefont {Majkov{\'a}}}, \bibinfo
  {author} {\bibfnamefont {E.}~\bibnamefont {Pinc{\'\i}k}}, \bibinfo {author}
  {\bibfnamefont {S.}~\bibnamefont {Luby}}, \bibinfo {author} {\bibfnamefont
  {L.}~\bibnamefont {Ortega}}, \bibinfo {author} {\bibfnamefont
  {R.}~\bibnamefont {Tucoulou}}, \bibinfo {author} {\bibfnamefont
  {P.}~\bibnamefont {Hudek}}, \ and\ \bibinfo {author} {\bibfnamefont
  {I.}~\bibnamefont {Kostic}},\ }\href {\doibase 10.1088/0022-3727/34/10A/339}
  {\bibfield  {journal} {\bibinfo  {journal} {J. Phys. D Appl. Phys.}\ }\textbf
  {\bibinfo {volume} {34}},\ \bibinfo {pages} {A188} (\bibinfo {year}
  {2001})}\BibitemShut {NoStop}%
\bibitem [{\citenamefont {Yan}\ \emph {et~al.}(2006)\citenamefont {Yan},
  \citenamefont {Bardeau}, \citenamefont {Brotons}, \citenamefont {Metzger},\
  and\ \citenamefont {Gibaud}}]{yan2006}%
  \BibitemOpen
  \bibfield  {author} {\bibinfo {author} {\bibfnamefont {M.}~\bibnamefont
  {Yan}}, \bibinfo {author} {\bibfnamefont {J.}~\bibnamefont {Bardeau}},
  \bibinfo {author} {\bibfnamefont {G.}~\bibnamefont {Brotons}}, \bibinfo
  {author} {\bibfnamefont {T.}~\bibnamefont {Metzger}}, \ and\ \bibinfo
  {author} {\bibfnamefont {A.}~\bibnamefont {Gibaud}},\ }in\ \href@noop {}
  {\emph {\bibinfo {booktitle} {KEK Proc.}}},\ \bibinfo {series and number}
  {2006-3}\ (\bibinfo {year} {2006})\ pp.\ \bibinfo {pages}
  {107--116}\BibitemShut {NoStop}%
\bibitem [{\citenamefont {Yan}\ and\ \citenamefont {Gibaud}(2007)}]{yan2007}%
  \BibitemOpen
  \bibfield  {author} {\bibinfo {author} {\bibfnamefont {M.}~\bibnamefont
  {Yan}}\ and\ \bibinfo {author} {\bibfnamefont {A.}~\bibnamefont {Gibaud}},\
  }\href {\doibase 10.1107/S0021889807044482} {\bibfield  {journal} {\bibinfo
  {journal} {J. Appl. Cryst.}\ }\textbf {\bibinfo {volume} {40}},\ \bibinfo
  {pages} {1050} (\bibinfo {year} {2007})}\BibitemShut {NoStop}%
\bibitem [{\citenamefont {Koch}, \citenamefont {Brown},\ and\ \citenamefont
  {Moncton}(1983)}]{koch_handbook_1983}%
  \BibitemOpen
  \bibinfo {editor} {\bibfnamefont {E.-E.}\ \bibnamefont {Koch}}, \bibinfo
  {editor} {\bibfnamefont {G.~S.}\ \bibnamefont {Brown}}, \ and\ \bibinfo
  {editor} {\bibfnamefont {D.~E.}\ \bibnamefont {Moncton}},\ eds.,\ \href@noop
  {} {\emph {\bibinfo {title} {Handbook on Synchrotron Radiation}}}\ (\bibinfo
  {publisher} {North Holland Publishing},\ \bibinfo {address} {Amsterdam ; New
  York},\ \bibinfo {year} {1983})\BibitemShut {NoStop}%
\bibitem [{\citenamefont {Naudon}\ \emph {et~al.}(2000)\citenamefont {Naudon},
  \citenamefont {Babonneau}, \citenamefont {Thiaudière},\ and\ \citenamefont
  {Lequien}}]{naudon_grazing-incidence_2000}%
  \BibitemOpen
  \bibfield  {author} {\bibinfo {author} {\bibfnamefont {A.}~\bibnamefont
  {Naudon}}, \bibinfo {author} {\bibfnamefont {D.}~\bibnamefont {Babonneau}},
  \bibinfo {author} {\bibfnamefont {D.}~\bibnamefont {Thiaudière}}, \ and\
  \bibinfo {author} {\bibfnamefont {S.}~\bibnamefont {Lequien}},\ }\href
  {\doibase 10.1016/S0921-4526(99)01894-3} {\bibfield  {journal} {\bibinfo
  {journal} {Physica B: Condensed Matter}\ }\textbf {\bibinfo {volume} {283}},\
  \bibinfo {pages} {69} (\bibinfo {year} {2000})}\BibitemShut {NoStop}%
\bibitem [{\citenamefont {Panduro}\ \emph {et~al.}(2014)\citenamefont
  {Panduro}, \citenamefont {Granlund}, \citenamefont {Sztucki}, \citenamefont
  {Konovalov}, \citenamefont {Breiby},\ and\ \citenamefont
  {Gibaud}}]{panduro_using_2014}%
  \BibitemOpen
  \bibfield  {author} {\bibinfo {author} {\bibfnamefont {E.~A.~C.}\
  \bibnamefont {Panduro}}, \bibinfo {author} {\bibfnamefont {H.}~\bibnamefont
  {Granlund}}, \bibinfo {author} {\bibfnamefont {M.}~\bibnamefont {Sztucki}},
  \bibinfo {author} {\bibfnamefont {O.}~\bibnamefont {Konovalov}}, \bibinfo
  {author} {\bibfnamefont {D.~W.}\ \bibnamefont {Breiby}}, \ and\ \bibinfo
  {author} {\bibfnamefont {A.}~\bibnamefont {Gibaud}},\ }\href {\doibase
  10.1021/am404602t} {\bibfield  {journal} {\bibinfo  {journal} {{ACS Appl.
  Mater. \& Interf.}}\ }\textbf {\bibinfo {volume} {6}},\ \bibinfo {pages}
  {2686} (\bibinfo {year} {2014})}\BibitemShut {NoStop}%
\bibitem [{\citenamefont {Cheng}\ \emph {et~al.}(2010)\citenamefont {Cheng},
  \citenamefont {Sanders}, \citenamefont {Truong}, \citenamefont {Harrer},
  \citenamefont {Friz}, \citenamefont {Holmes}, \citenamefont {Colburn},\ and\
  \citenamefont {Hinsberg}}]{cheng2010fabri}%
  \BibitemOpen
  \bibfield  {author} {\bibinfo {author} {\bibfnamefont {J.}~\bibnamefont
  {Cheng}}, \bibinfo {author} {\bibfnamefont {D.}~\bibnamefont {Sanders}},
  \bibinfo {author} {\bibfnamefont {H.}~\bibnamefont {Truong}}, \bibinfo
  {author} {\bibfnamefont {S.}~\bibnamefont {Harrer}}, \bibinfo {author}
  {\bibfnamefont {A.}~\bibnamefont {Friz}}, \bibinfo {author} {\bibfnamefont
  {S.}~\bibnamefont {Holmes}}, \bibinfo {author} {\bibfnamefont
  {M.}~\bibnamefont {Colburn}}, \ and\ \bibinfo {author} {\bibfnamefont
  {W.}~\bibnamefont {Hinsberg}},\ }\href {\doibase 10.1021/nn100686v}
  {\bibfield  {journal} {\bibinfo  {journal} {{ACS} Nano}\ }\textbf {\bibinfo
  {volume} {4}},\ \bibinfo {pages} {4815} (\bibinfo {year} {2010})}\BibitemShut
  {NoStop}%
\bibitem [{\citenamefont {Krumrey}\ and\ \citenamefont
  {Ulm}(2001)}]{krumrey2001}%
  \BibitemOpen
  \bibfield  {author} {\bibinfo {author} {\bibfnamefont {M.}~\bibnamefont
  {Krumrey}}\ and\ \bibinfo {author} {\bibfnamefont {G.}~\bibnamefont {Ulm}},\
  }\href {\doibase 10.1016/S0168-9002(01)00598-8} {\bibfield  {journal}
  {\bibinfo  {journal} {Nucl. Instr. Meth. A}\ }\textbf {\bibinfo {volume}
  {467-468}},\ \bibinfo {pages} {1175} (\bibinfo {year} {2001})}\BibitemShut
  {NoStop}%
\bibitem [{\citenamefont {Beckhoff}\ \emph {et~al.}(2009)\citenamefont
  {Beckhoff}, \citenamefont {Gottwald}, \citenamefont {Klein}, \citenamefont
  {Krumrey}, \citenamefont {M{\"u}ller}, \citenamefont {Richter}, \citenamefont
  {Scholze}, \citenamefont {Thornagel},\ and\ \citenamefont
  {Ulm}}]{ptb-quarterc}%
  \BibitemOpen
  \bibfield  {author} {\bibinfo {author} {\bibfnamefont {B.}~\bibnamefont
  {Beckhoff}}, \bibinfo {author} {\bibfnamefont {A.}~\bibnamefont {Gottwald}},
  \bibinfo {author} {\bibfnamefont {R.}~\bibnamefont {Klein}}, \bibinfo
  {author} {\bibfnamefont {M.}~\bibnamefont {Krumrey}}, \bibinfo {author}
  {\bibfnamefont {R.}~\bibnamefont {M{\"u}ller}}, \bibinfo {author}
  {\bibfnamefont {M.}~\bibnamefont {Richter}}, \bibinfo {author} {\bibfnamefont
  {F.}~\bibnamefont {Scholze}}, \bibinfo {author} {\bibfnamefont
  {R.}~\bibnamefont {Thornagel}}, \ and\ \bibinfo {author} {\bibfnamefont
  {G.}~\bibnamefont {Ulm}},\ }\href@noop {} {\bibfield  {journal} {\bibinfo
  {journal} {phys. status solidi b}\ }\textbf {\bibinfo {volume} {246}},\
  \bibinfo {pages} {1415} (\bibinfo {year} {2009})}\BibitemShut {NoStop}%
\bibitem [{\citenamefont {Fuchs}\ \emph {et~al.}(1995)\citenamefont {Fuchs},
  \citenamefont {Krumrey}, \citenamefont {M{\"u}ller}, \citenamefont
  {Scholze},\ and\ \citenamefont {Ulm}}]{fuchs-newref}%
  \BibitemOpen
  \bibfield  {author} {\bibinfo {author} {\bibfnamefont {D.}~\bibnamefont
  {Fuchs}}, \bibinfo {author} {\bibfnamefont {M.}~\bibnamefont {Krumrey}},
  \bibinfo {author} {\bibfnamefont {P.}~\bibnamefont {M{\"u}ller}}, \bibinfo
  {author} {\bibfnamefont {F.}~\bibnamefont {Scholze}}, \ and\ \bibinfo
  {author} {\bibfnamefont {G.}~\bibnamefont {Ulm}},\ }\href@noop {} {\bibfield
  {journal} {\bibinfo  {journal} {Rev. Sci. Instrum.}\ }\textbf {\bibinfo
  {volume} {66}},\ \bibinfo {pages} {2248} (\bibinfo {year}
  {1995})}\BibitemShut {NoStop}%
\bibitem [{\citenamefont {Hoell}\ \emph {et~al.}(2007)\citenamefont {Hoell},
  \citenamefont {Zizak}, \citenamefont {Bieder},\ and\ \citenamefont
  {Mokrani}}]{hoell-saxs-patent}%
  \BibitemOpen
  \bibfield  {author} {\bibinfo {author} {\bibfnamefont {A.}~\bibnamefont
  {Hoell}}, \bibinfo {author} {\bibfnamefont {I.}~\bibnamefont {Zizak}},
  \bibinfo {author} {\bibfnamefont {H.}~\bibnamefont {Bieder}}, \ and\ \bibinfo
  {author} {\bibfnamefont {L.}~\bibnamefont {Mokrani}},\ }\href@noop {}
  {\enquote {\bibinfo {title} {German patent {DE} 10 2006 029 449},}\ }
  (\bibinfo {year} {2007})\BibitemShut {NoStop}%
\bibitem [{\citenamefont {{{BIPM}}}\ \emph {et~al.}(2008)\citenamefont
  {{{BIPM}}}, \citenamefont {{{IEC}}}, \citenamefont {{{IFCC}}}, \citenamefont
  {{{ILAC}}}, \citenamefont {{{IUPAC}}}, \citenamefont {{{IUPAP}}},
  \citenamefont {{{ISO}}},\ and\ \citenamefont {{{OIML}}}}]{gum}%
  \BibitemOpen
  \bibfield  {author} {\bibinfo {author} {\bibnamefont {{{BIPM}}}}, \bibinfo
  {author} {\bibnamefont {{{IEC}}}}, \bibinfo {author} {\bibnamefont
  {{{IFCC}}}}, \bibinfo {author} {\bibnamefont {{{ILAC}}}}, \bibinfo {author}
  {\bibnamefont {{{IUPAC}}}}, \bibinfo {author} {\bibnamefont {{{IUPAP}}}},
  \bibinfo {author} {\bibnamefont {{{ISO}}}}, \ and\ \bibinfo {author}
  {\bibnamefont {{{OIML}}}},\ }\href@noop {} {\emph {\bibinfo {title}
  {{Evaluation of measurement data - Guide for the expression of uncertainty in
  measurement. JCGM 100: 2008}}}}\ (\bibinfo  {publisher} {{BIPM} Joint
  Committee for Guides in Metrology, Paris, S\`{e}vres},\ \bibinfo {year}
  {2008})\ \Eprint
  {http://arxiv.org/abs/http://www.bipm.org/en/publications/guides/gum.html}
  {http://www.bipm.org/en/publications/guides/gum.html} \BibitemShut {NoStop}%
\bibitem [{\citenamefont {Wernecke}\ \emph {et~al.}(2014)\citenamefont
  {Wernecke}, \citenamefont {Gollwitzer}, \citenamefont {M\"{u}ller},\ and\
  \citenamefont {Krumrey}}]{invac-pilatus}%
  \BibitemOpen
  \bibfield  {author} {\bibinfo {author} {\bibfnamefont {J.}~\bibnamefont
  {Wernecke}}, \bibinfo {author} {\bibfnamefont {C.}~\bibnamefont
  {Gollwitzer}}, \bibinfo {author} {\bibfnamefont {P.}~\bibnamefont
  {M\"{u}ller}}, \ and\ \bibinfo {author} {\bibfnamefont {M.}~\bibnamefont
  {Krumrey}},\ }\href {\doibase 10.1107/S160057751400294X} {\bibfield
  {journal} {\bibinfo  {journal} {J. Synchrotron Radiat.}\ }\textbf {\bibinfo
  {volume} {21}},\ \bibinfo {pages} {529} (\bibinfo {year} {2014})}\BibitemShut
  {NoStop}%
\end{thebibliography}%

\end{document}